\documentclass[prd,aps,twocolumn,superscriptaddress,nofootinbib,longbibliography,10pt]{revtex4-2}

\usepackage{amsmath,amssymb}
\usepackage{graphicx}
\usepackage{enumitem}

\newcommand{\order}{\mathcal{O}}
\newcommand{\emdash}{\,---\,}

\begin{document}

\title{Electromagnetic, gravitational wave, and static gravitational transmission through throat spacetimes: a constraint-wave asymmetry}

\author{Jeff Riley}
\affiliation{School of Physics and Astronomy, Monash University, Clayton, Victoria 3800, Australia}
\affiliation{OzGrav, ARC Centre of Excellence for Gravitational Wave Discovery, Australia}
\email{jeff.riley@monash.edu}

%\date{\today}

\begin{abstract}
We compute the transmission properties of electromagnetic (EM), gravitational wave (GW), and static gravitational perturbations through geometric throats in spherically symmetric spacetimes. On the ultrastatic Ellis-Bronnikov background, decomposition of the four-dimensional Maxwell equations into vector spherical harmonics yields an effective Schr\"odinger problem with centrifugal barrier $V_\ell^{(\mathrm{EM})}=\ell(\ell+1)/(\sigma^2+r_0^2)$ peaked at the throat. For the lowest physical EM mode ($\ell=1$), frequencies below the barrier-top frequency $\omega_{\max}=\sqrt{2}/r_0$ are strongly suppressed by sub-barrier tunnelling. Gravitational wave perturbations ($\ell\ge 2$) see a qualitatively similar barrier and are likewise strongly suppressed below their respective barrier-top frequencies. By contrast, the static gravitational monopole ($\ell=0$), governed by the linearised Einstein equations on the same background, satisfies the source-free conservation law $(a^2\Phi')'=0$ with no potential barrier, yielding the exact solution $\Phi\propto\arctan(\sigma/r_0)$. We extend these results to a one-parameter family of throat geometries with varying profile shapes, and to a reflected-Schwarzschild (Damour-Solodukhin-type) wormhole, demonstrating that the qualitative asymmetry\emdash strong sub-barrier suppression for all propagating radiation ($\ell\ge 1$) versus polynomial attenuation for the static monopole ($\ell=0$)\emdash is universal for static, spherically symmetric throats. Numerov integration, WKB estimates, and exact analytical solutions are compared throughout. The results establish a structural constraint-wave asymmetry arising from the multipole decomposition of the field equations, independent of the matter content sourcing the geometry, on a fixed background.
\end{abstract}

\maketitle

%====================================================================%
\section{Introduction}
\label{sec:intro}

The propagation of fields through spacetime geometries containing throats\emdash minimal-area two-spheres connecting two asymptotic regions\emdash has been studied extensively in the context of wormhole physics~\cite{MorrisThorne1988,Visser1995,Bronnikov2018}. Since the foundational work of Morris and Thorne~\cite{MorrisThorne1988}, traversable wormholes have served as a theoretical laboratory for understanding field propagation in non-trivial topologies. Scattering calculations on wormhole backgrounds have determined reflection and transmission coefficients for scalar~\cite{Konoplya2005,Aneesh2018}, electromagnetic~\cite{KonoplyaZhidenko2010,Churilova2021}, and gravitational perturbations~\cite{KimSung2008,KonoplyaZhidenko2010}, with applications to quasinormal mode spectra, late-time tails, and echoes as potential observational signatures~\cite{CardosoPani2019,BuenoEtAl2018,DamourSolodukhin2007}. Dai and Stojkovic~\cite{DaiStojkovic2019} demonstrated that the gravitational monopole of a source on one side of a traversable wormhole is detectable on the other side, with higher static multipoles suppressed by geometric factors \mbox{$(R/A)^{2\ell+1}$}, and proposed observational tests based on stellar orbits near Sgr~A*. Our analysis differs in treating frequency-dependent wave propagation on smooth throat backgrounds rather than static potential matching on thin-shell geometries, but the monopole transmission they identify is the static limit of the constraint-wave asymmetry we derive here.

The effective potentials governing these scattering problems are well understood: for any static, spherically symmetric spacetime, the angular decomposition of massless field equations yields a one-dimensional Schr\"odinger-type equation with an effective potential determined by the background geometry~\cite{Chandrasekhar1983,ReggeWheeler1957,Zerilli1970}. On wormhole backgrounds, these potentials typically peak at or near the throat, creating a barrier whose height and width determine the scattering properties~\cite{KimSung2008,KonoplyaZhidenko2010}. The extensive literature on wormhole perturbation theory has largely focused on the above-barrier regime (quasinormal modes, scattering cross sections) or on the spectral properties of the potential (quasi-bound states, echoes)~\cite{CardosoPani2019,MarkEtAl2017,BuenoEtAl2018}.

A feature of these scattering problems that has received comparatively less attention is the structural asymmetry between different field sectors in the sub-barrier regime. For monochromatic waves with frequency below the peak of the effective potential, EM multipoles ($\ell\ge 1$) and gravitational wave (GW) multipoles ($\ell\ge 2$) encounter a centrifugal barrier and are strongly attenuated by sub-barrier tunnelling, in direct analogy with barrier penetration in non-relativistic quantum mechanics~\cite{LandauLifshitz_QM}. Rigorous transmission bounds for such barriers have been derived by Boonserm and Visser~\cite{BoonsermVisser2009}. The static gravitational monopole ($\ell=0$), by contrast, satisfies a Poisson-type (elliptic) equation whose $\ell=0$ sector has no centrifugal term and therefore no barrier. The monopole transmits through the throat with only polynomial (geometric) attenuation.

This distinction is a consequence of the fundamental difference between hyperbolic (wave) and elliptic (Poisson/constraint) equations\emdash the former admits oscillatory and evanescent regimes separated by a potential barrier, while the latter does not. The barrier for propagating radiation has the familiar centrifugal form $V_\ell\propto\ell(\ell+1)/a^2$, where $a$ is the areal radius\emdash this is positive for all $\ell\ge 1$ and vanishes identically for $\ell=0$. Although the individual ingredients\emdash effective potentials on wormhole backgrounds~\cite{Chandrasekhar1983,KimSung2008}, monopole solutions of the Laplace equation~\cite{Visser1995}\emdash are well known, a systematic quantitative comparison of the transmission properties across all three sectors (EM, GW, static gravity) on the same background, covering the full sub-barrier frequency range, and extending across different throat geometries, has not been presented.

We provide such a comparison in this paper. Working primarily on the Ellis-Bronnikov (EB) ultrastatic throat~\cite{Ellis1973,Bronnikov1973}, and extending to a parametric family of throat profiles and the Damour-Solodukhin wormhole~\cite{DamourSolodukhin2007}, we:
\begin{enumerate}[leftmargin=2em,label=(\roman*)]

\item derive the effective potentials for EM and GW perturbations from the four-dimensional field equations via vector and tensor spherical harmonic decomposition (Sec.~\ref{sec:decomposition});

\item compute the EM and GW transmission coefficients $T(\omega)$ by Numerov integration and compare with WKB estimates (Sections~\ref{sec:EM} and~\ref{sec:GW});

\item solve the linearised Einstein equations for the static gravitational monopole and obtain the exact solution, and show that higher ($\ell\ge 1$) gravitational multipoles see barriers analogous to EM (Sec.~\ref{sec:monopole});

\item extend the results to a family of throat profiles and the Damour-Solodukhin wormhole, demonstrating universality of the asymmetry (Sec.~\ref{sec:universality});

\item quantify the constraint-wave asymmetry and discuss its physical origin and implications (Sec.~\ref{sec:asymmetry}).

\end{enumerate}

Unless otherwise stated we use units with $c=G=1$, and metric signature $(-,+,+,+)$. We denote the proper radial coordinate by~$\sigma$, and reserve~$\ell$ exclusively for the angular momentum multipole index, to avoid ambiguity.

%====================================================================%
\section{Throat geometries}
\label{sec:geometry}

\subsection{General static spherically symmetric metric}

We consider static, spherically symmetric spacetimes of the form
\begin{equation}
ds^2 = -e^{2\alpha(\sigma)}\,dt^2 + d\sigma^2 + a(\sigma)^2\,d\Omega^2,
\label{eq:general_metric}
\end{equation}
where $\sigma\in(-\infty,+\infty)$ is the proper radial coordinate, $d\Omega^2=d\theta^2+\sin^2\!\theta\,d\varphi^2$, and the areal radius $a(\sigma)$ has a local minimum at $\sigma=0$ (the ``throat'') with \mbox{$a(0)=r_0>0$}. The function $\alpha(\sigma)$ determines the gravitational redshift. A spacetime of this form has a throat provided \mbox{$a'(0)=0$ and $a''(0)>0$}. The area $4\pi a(\sigma)^2$ of the $(\theta,\varphi)$ two-spheres has a minimum $4\pi r_0^2$ at the throat. The areal radius~$a$ and the effective potentials derived below are coordinate-invariant geometric quantities.

\subsection{Ellis-Bronnikov throat}

The Ellis-Bronnikov (EB) spacetime~\cite{Ellis1973,Bronnikov1973} is the simplest ultrastatic ($\alpha=0$) wormhole:
\begin{equation}
ds^2 = -dt^2 + d\sigma^2 + (\sigma^2+r_0^2)\,d\Omega^2.
\label{eq:EB}
\end{equation}
The areal radius is $a(\sigma)=\sqrt{\sigma^2+r_0^2}$, giving a smooth throat at $\sigma=0$. This spacetime is sourced by a phantom (negative kinetic energy) scalar field and violates the null energy condition~\cite{MorrisThorne1988,Visser1995}. For the purpose of this paper, which concerns field propagation on a fixed background, the nature of the source is immaterial\emdash the effective potentials and transmission coefficients depend only on the geometry.

The spatial Ricci scalar of the $t=\mathrm{const}$ slices is
\begin{equation}
R^{(3)} = -\frac{2r_0^2}{(\sigma^2+r_0^2)^2}\,,
\label{eq:R3_EB}
\end{equation}
which is everywhere negative, with minimum \mbox{$R^{(3)}(0)=-2/r_0^2$}.

\subsection{Parametric throat family}

\begin{figure*}[!t]
\centering
\includegraphics[width=\textwidth]{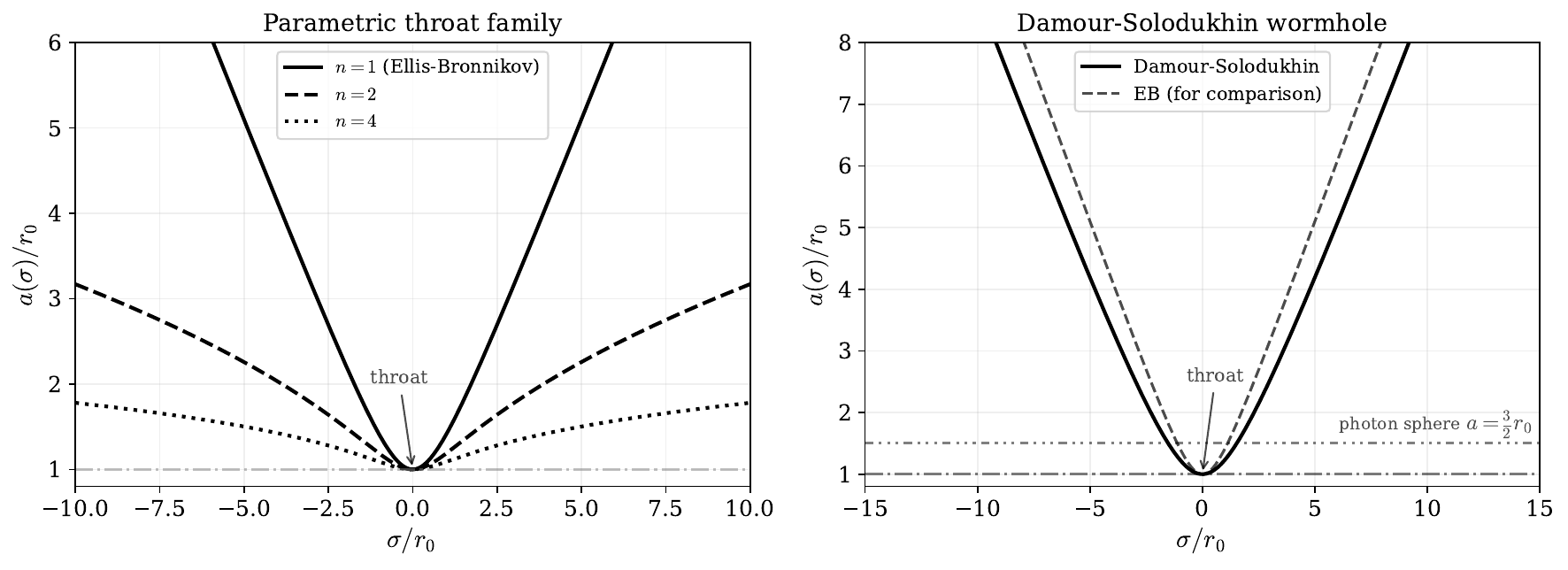}
\caption{Areal radius profiles $a(\sigma)/r_0$ for the throat geometries studied in this paper. Left panel:~the parametric family~\eqref{eq:parametric}. $n=1$ is the Ellis-Bronnikov profile\emdash increasing $n$ produces a broader, flatter throat with the same minimum radius~$r_0$. Right panel:~the Damour-Solodukhin reflected-Schwarzschild wormhole (solid), compared with EB (dashed). The DS profile rises more gradually near the throat than EB, reflecting the vanishing lapse ($e^{2\alpha}\to 0$) at $\sigma=0$. The photon sphere at $a=3r_0/2$ is shown.}
\label{fig:profiles}
\end{figure*}

To study how the barrier shape affects suppression, we introduce a one-parameter family of throat profiles:
\begin{equation}
a_n(\sigma) = r_0\!\left(1+\frac{\sigma^2}{r_0^2}\right)^{1/(2n)}\!\!,\quad n=1,2,3,\ldots
\label{eq:parametric}
\end{equation}
For $n=1$, this reduces to the EB profile. As $n$ increases, the throat becomes flatter (more plateau-like) near $\sigma=0$, while maintaining the same minimum $a_n(0)=r_0$ (Fig.~\ref{fig:profiles}, left panel). The barrier height $V_\ell(0)=\ell(\ell+1)/r_0^2$ is $n$-independent, but the barrier width increases with~$n$, enhancing the tunnelling suppression. The $n\to\infty$ limit of~\eqref{eq:parametric} gives $a_\infty(\sigma)=r_0$ for all~$\sigma$, i.e., an infinite cylinder of constant radius, which is not asymptotically flat. For finite~$n$, the geometry is asymptotically flat and transmission is well defined via plane-wave boundary conditions at $|\sigma|\to\infty$. The $n\to\infty$ limit is a conceptual extrapolation; operationally, the values in Table~\ref{tab:parametric} for large~$n$ are computed on a finite domain $|\sigma|\le L_{\max}$ with $L_{\max}\gg r_0$ and are insensitive to $L_{\max}$ provided $L_{\max}$ exceeds the region where $a_n$ differs appreciably from~$r_0$. A finite-length cylindrical neck (radius~$r_0$, length~$L$, joined to flat geometry at $|\sigma|=L/2$) would be a separate model producing compact-support exponential suppression $T\sim\exp(-2\sqrt{\ell(\ell+1)}\,L/r_0)$\emdash the parametric family~\eqref{eq:parametric} does not encode such a finite neck.

\subsection{Damour-Solodukhin wormhole}

As a non-ultrastatic test case, we consider a reflected-Schwarzschild wormhole inspired by the construction of Damour and Solodukhin~\cite{DamourSolodukhin2007}. The Schwarzschild exterior metric for $a>r_0$,
\begin{equation}
ds^2 = -\!\left(1-\frac{r_0}{a}\right)dt^2 + \frac{da^2}{1-r_0/a} + a^2\,d\Omega^2,
\label{eq:DS}
\end{equation}
is reflected across $a=r_0$ to create a second asymptotic region, producing a traversable throat (a minimal-area surface) where $g_{tt}\to 0$. The original DS parameterisation~\cite{DamourSolodukhin2007} introduces a small deformation parameter, $\lambda$, that keeps $g_{tt}>0$ at the throat\emdash our construction corresponds to the $\lambda\to 0$ limit. In terms of the proper radial coordinate~$\sigma$, the metric takes the form~\eqref{eq:general_metric}, with $e^{2\alpha}=1-r_0/a(\sigma)$. This geometry has a ``photon sphere'' at $a=3r_0/2$ and a qualitatively different (double-peaked) barrier structure (see Fig.~\ref{fig:DS} in Section~\ref{sec:DS_wormhole_analysis}).

%====================================================================%
\section{Spherical harmonic decomposition}
\label{sec:decomposition}

\subsection{Electromagnetic perturbations}

Source-free Maxwell's equations $\nabla_\mu F^{\mu\nu}=0$ on the background~\eqref{eq:general_metric} are decomposed into two independent sectors using vector spherical harmonics~\cite{Chandrasekhar1983,ReggeWheeler1957}.

The axial (magnetic) gauge-invariant master variable $\Psi_\ell^{(B)}$ satisfies
\begin{equation}
-\frac{\partial^2\Psi_\ell^{(B)}}{\partial t^2} + \frac{\partial^2\Psi_\ell^{(B)}}{\partial\sigma_*^2} - V_\ell^{(\mathrm{EM})}\,\Psi_\ell^{(B)} = 0,
\label{eq:axial_EM}
\end{equation}
where $\sigma_*$ is the tortoise coordinate defined by \mbox{$d\sigma_*=e^{-\alpha}\,d\sigma$}, and the effective potential is
\begin{equation}
V_\ell^{(\mathrm{EM})} = e^{2\alpha}\frac{\ell(\ell+1)}{a^2}\,.
\label{eq:V_EM_general}
\end{equation}
The polar (electric) master variable satisfies the same equation with the same potential\emdash a well-known isospectrality of Maxwell's equations on spherically symmetric backgrounds~\cite{Chandrasekhar1983}.

For the ultrastatic EB metric ($\alpha=0$, $\sigma_*=\sigma$), the potential reduces to
\begin{equation}
V_\ell^{(\mathrm{EM})} = \frac{\ell(\ell+1)}{\sigma^2+r_0^2}\,,
\label{eq:V_EM_EB}
\end{equation}
with maximum $V_\ell(0)=\ell(\ell+1)/r_0^2$ at the throat. The barrier-top frequency\footnote{We use ``barrier-top frequency'' for $\omega_{\max}=\sqrt{V_\ell(0)}$ to avoid confusion with a strict cutoff: the transmission coefficient is nonzero for all $\omega>0$, decreasing continuously below $\omega_{\max}$ rather than vanishing abruptly. In the deep sub-barrier regime, however, the suppression is so strong that $\omega_{\max}$ plays a role analogous to a waveguide cutoff frequency.} for the lowest physical EM mode ($\ell=1$) is
\begin{equation}
\omega_{\max}^{(\ell=1)} = \frac{\sqrt{2}}{r_0}\,.
\label{eq:cutoff_EM}
\end{equation}

\subsection{Gravitational wave perturbations}

Gravitational perturbations decompose into axial (Regge-Wheeler) and polar (Zerilli) sectors~\cite{ReggeWheeler1957,Zerilli1970,Chandrasekhar1983}. For the EB background, the axial sector for $\ell\ge 2$ yields
\begin{equation}
\frac{d^2\Psi_\ell^{(\mathrm{GW})}}{d\sigma^2} + \left[\omega^2 - V_\ell^{(\mathrm{GW})}(\sigma)\right]\Psi_\ell^{(\mathrm{GW})} = 0,
\label{eq:GW_eqn}
\end{equation}
with effective potential
\begin{equation}
V_\ell^{(\mathrm{GW})} = \frac{\ell(\ell+1)}{\sigma^2+r_0^2} - \frac{3r_0^2}{(\sigma^2+r_0^2)^2}\,.
\label{eq:V_GW_EB}
\end{equation}
The first term is the centrifugal barrier (identical to the EM case), while the second is a curvature correction, $-3r_0^2/a^4$, arising from the non-flat spatial Ricci tensor of the EB background~\cite{KimSung2008}. For $\ell=2$, the barrier height is $V_2^{(\mathrm{GW})}(0)=3/r_0^2$, compared with $V_2^{(\mathrm{EM})}(0)=6/r_0^2$\emdash the GW barrier is lower than the EM barrier for the same multipole. The GW barrier-top frequency for $\ell=2$ is $\omega_{\max}^{(\mathrm{GW},\ell=2)}=\sqrt{3}/r_0\approx 1.73/r_0$.

\subsection{Static gravitational monopole equation}

The static gravitational potential in the weak-field limit is governed by the Poisson equation on the spatial background. From the ADM formalism with $K_{ij}=0$, the trace of the lapse equation gives~\cite{MTW1973,Wald1984}
\begin{equation}
\nabla^2\Phi = 4\pi(\rho+S),
\label{eq:poisson}
\end{equation}
where $\nabla^2$ is the covariant Laplacian of the spatial metric and no additional curvature coupling appears~\cite{Wald1984}.

For the ultrastatic case ($\alpha=0$), the $\ell=0$ component on the spatial metric $d\sigma^2+a(\sigma)^2\,d\Omega^2$ yields the source-free equation
\begin{equation}
\frac{1}{a^2}\frac{d}{d\sigma}\!\left(a^2\frac{d\Phi}{d\sigma}\right) = 0.
\label{eq:monopole_ultrastatic}
\end{equation}
This is a conservation law\emdash the flux $\mathcal{F}=a^2\Phi'$ is constant. There is no centrifugal barrier and no sub-barrier suppression, for any throat profile~$a(\sigma)$. The generalisation to non-ultrastatic backgrounds ($\alpha\neq 0$), where the conserved flux becomes $\mathcal{F}=a^2 e^{2\alpha}\Phi'$, is derived in Appendix~\ref{app:monopole}.

%====================================================================%
\section{Electromagnetic transmission}
\label{sec:EM}

\subsection{WKB estimate and low-frequency scaling}

For $\omega<\omega_{\max}^{(\ell)}$, the classical turning points $\pm\sigma_{\mathrm{tp}}$ satisfy \mbox{$V_\ell(\sigma_{\mathrm{tp}})=\omega^2$}, giving \mbox{$\sigma_{\mathrm{tp}}=\sqrt{\ell(\ell+1)/\omega^2-r_0^2}$} on the EB throat. The WKB transmission coefficient is
\begin{equation}
T_{\mathrm{WKB}} = \frac{1}{1+\exp\!\left(2\!\int_{-\sigma_{\mathrm{tp}}}^{+\sigma_{\mathrm{tp}}}\!\!\sqrt{V_\ell(\sigma)-\omega^2}\;d\sigma\right)}.
\label{eq:WKB}
\end{equation}
This gives $T=1/2$ at the barrier summit ($S=0$) and reduces to $T\approx\exp(-2S)$ deep in the sub-barrier regime ($S\gg 1$). In the low-frequency limit, the turning points recede as \mbox{$\sigma_{\mathrm{tp}}\approx\sqrt{\ell(\ell+1)}/\omega\to\infty$}.

Because the EB potential decays as \mbox{$V_\ell\sim\ell(\ell+1)/\sigma^2$} at large~$|\sigma|$ (a long-range centrifugal tail), the WKB action grows logarithmically with $1/\omega$, producing power-law suppression:
\begin{equation}
T(\omega) \sim K\,(\omega r_0)^\nu, \quad \omega r_0\ll 1,
\label{eq:TWKB_powerlaw}
\end{equation}
where $K$ and $\nu$ depend on the barrier shape. Standard low-energy scattering theory for a $1/\sigma^2$ centrifugal tail predicts that the power transmission coefficient scales as $T\propto\omega^{2\ell+1}$ (i.e., $\nu=2\ell+1$) in the strict $\omega\to 0$ limit~\cite{LandauLifshitz_QM,Newton1982}. The power-law character of the sub-barrier suppression contrasts with barriers of compact support (finite spatial extent), which produce exponential suppression \mbox{$T\sim e^{-2\gamma L/r_0}$}, with \mbox{$\gamma=\sqrt{\ell(\ell+1)}$}.

\subsection{Numerical solution}

We solve Eq.~\eqref{eq:axial_EM} numerically using Numerov's method on a uniform grid $\sigma\in[-L_{\max},+L_{\max}]$, with \mbox{$L_{\max}=30\,r_0$} and \mbox{$N=10^4$} points. The boundary condition at \mbox{$\sigma=+L_{\max}$} is a unit-amplitude incident plane wave \mbox{$\Psi=e^{-i\omega\sigma}$}. The transmitted amplitude at \mbox{$\sigma=-L_{\max}$} is extracted by decomposing the solution into left- and right-moving components.

For the EB potential with its $1/\sigma^2$ tail, the plane-wave decomposition is valid provided $\omega L_{\max}\gg 1$, which holds for all frequencies in Table~\ref{tab:TEM}. The impact of using exact Bessel asymptotics instead of plane waves is discussed in Appendix~\ref{app:numerics}. Convergence is verified by doubling $N$ and $L_{\max}$.

For deep sub-barrier frequencies ($\omega r_0\le 0.1$), numerical precision is the main concern — the transmitted amplitude is $|A|\gg 1$ and the transmission $T=|A|^{-2}$ involves cancellation between large numbers. We work in standard double precision ($\sim\!15$ significant digits) and verify that $T$ is stable to at least two significant figures under doubling of $N$ for all tabulated values with $T\ge 10^{-15}$. For $T < 10^{-15}$, the extraction becomes unreliable in double precision and these values should be interpreted as upper bounds.

\begin{figure*}[!tbp]
\centering
\includegraphics[width=0.65\textwidth]{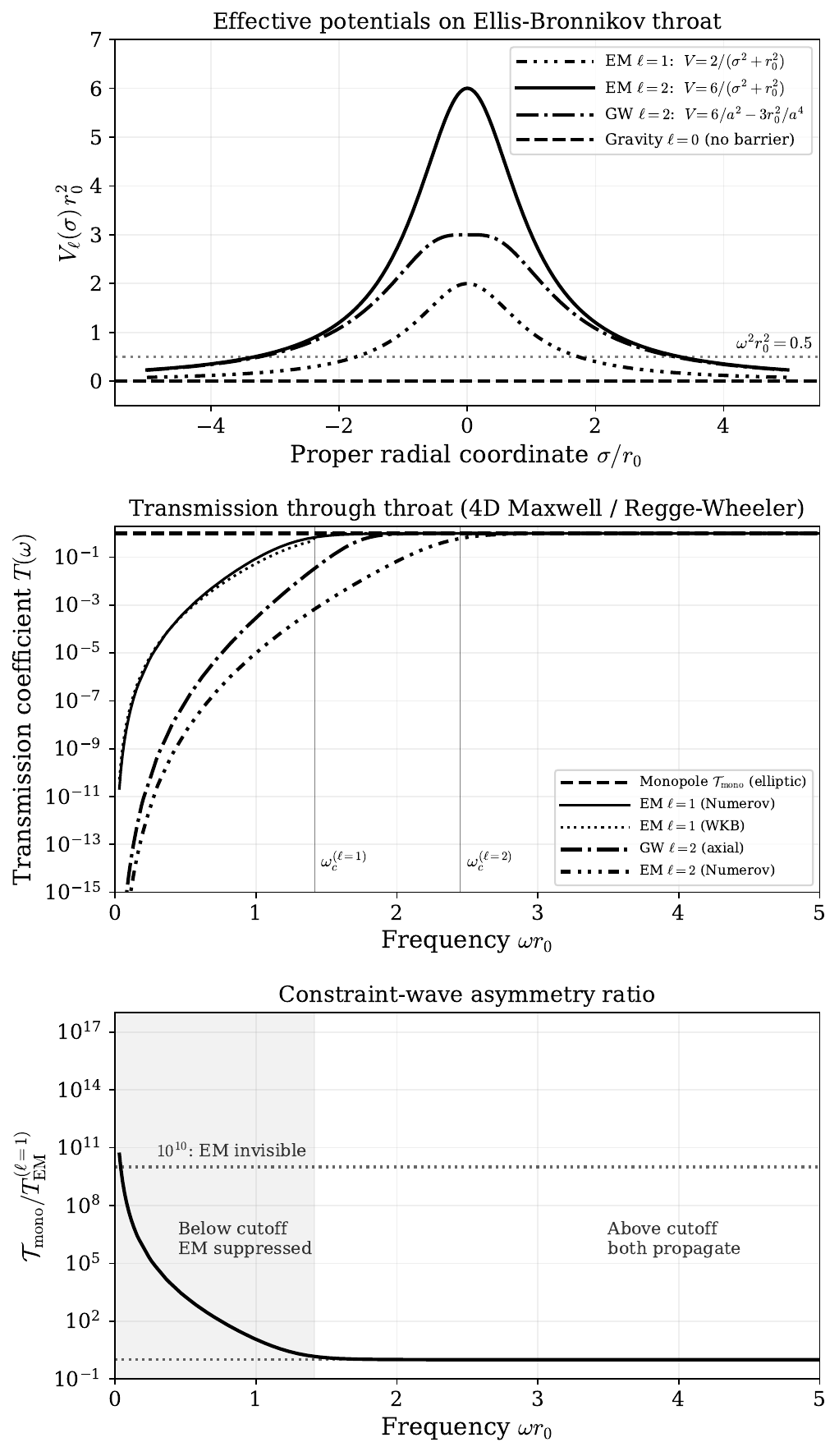}
\caption{Constraint-wave asymmetry on the Ellis-Bronnikov throat ($r_0=1$).\\
Upper panel:~Effective potentials $V_\ell(\sigma)$ derived from the 4D field equations, plotted against proper radial coordinate $\sigma$.
EM modes at $\ell=1$ (dots-dash) and $\ell=2$ (solid), and the GW axial mode at $\ell=2$ (dash-dot), see centrifugal barriers peaked at the throat.
The curvature correction reduces the GW $\ell=2$ barrier height to $3/r_0^2$, half the EM $\ell=2$ value of $6/r_0^2$ (Table~\ref{tab:EMvsGW_l2}).
The $\ell=0$ gravitational monopole (dashed horizontal line at $V=0$) satisfies the first-order conservation law $(a^2\Phi')'=0$, which does not admit a Schr\"odinger-form effective potential and sees no centrifugal barrier.\\
Centre panel:~Power transmission coefficients $T(\omega)$: EM and GW modes are strongly suppressed below their respective barrier-top frequencies; the monopole reference line shows $\mathcal{T}_{\mathrm{mono}}$ (Eq.~\eqref{eq:T_mono}), which is an elliptic flux ratio and therefore $\omega$-independent.\\
Lower panel:~The ratio $\mathcal{T}_{\mathrm{mono}}/T_{\mathrm{EM}}^{(\ell=1)}$ illustrates the growing asymmetry at low frequencies.
Note that these are conceptually distinct quantities\emdash static constraint versus wave scattering\emdash and their ratio should be interpreted as a measure of the asymmetry, not as a transmission coefficient.}
\label{fig:scattering}
\end{figure*}

The upper panel of Figure~\ref{fig:scattering} shows the effective potentials for the EM ($\ell=1,2$), GW ($\ell=2$), and monopole ($\ell=0$) sectors on the EB throat. The centre panel of Figure~\ref{fig:scattering} shows the transmission coefficients. Table~\ref{tab:TEM} presents the EM results at selected frequencies. The Numerov results confirm strong sub-barrier suppression for all $\ell\ge 1$ modes, in qualitative agreement with the WKB estimates.

\begin{table}[b]
\caption{Transmission coefficients on the Ellis-Bronnikov throat ($r_0=1$). EM barrier-top frequencies: \mbox{$\omega_{\max}^{(\ell=1)}=\sqrt{2}/r_0\approx 1.414/r_0$, and $\omega_{\max}^{(\ell=2)}=\sqrt{6}/r_0\approx 2.449/r_0$.}}
\label{tab:TEM}
\begin{ruledtabular}
\begin{tabular}{cccc}
$\omega r_0$ & $T_{\mathrm{EM}}^{(\ell=1)}$ & $T_{\mathrm{WKB}}^{(\ell=1)}$ & $T_{\mathrm{EM}}^{(\ell=2)}$ \\
\hline
0.1 & $2.2\times 10^{-8}$ & $3.6\times 10^{-8}$ & $<10^{-15}$ \\
0.3 & $1.8\times 10^{-5}$ & $2.1\times 10^{-5}$ & $<10^{-12}$ \\
0.5 & $5.4\times 10^{-4}$ & $4.9\times 10^{-4}$ & $4.9\times 10^{-9}$ \\
0.7 & $5.3\times 10^{-3}$ & $4.0\times 10^{-3}$ & $4.5\times 10^{-6}$ \\
1.0 & $9.5\times 10^{-2}$ & $6.3\times 10^{-2}$ & $1.1\times 10^{-4}$ \\
$\sqrt{2}$ & $0.70$ & 0.50 & $1.3\times 10^{-2}$ \\
2.0 & $0.99$ & $1.00$ & $0.24$ \\
3.0 & $1.00$ & $1.00$ & $0.99$ \\
\end{tabular}
\end{ruledtabular}
\end{table}

\begin{figure}[b]
\centering
\includegraphics[width=\columnwidth]{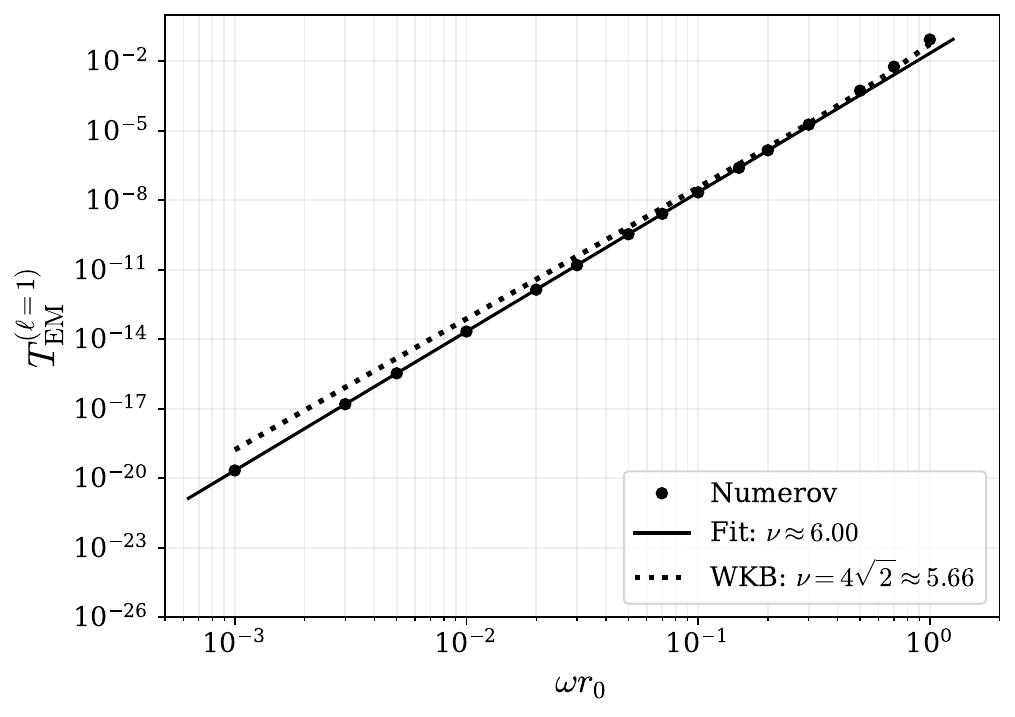}
\caption{Low-frequency power-law scaling of the EM $\ell=1$ transmission coefficient on the EB throat.
Markers:~ Numerov integration; solid line:~fit \mbox{$T=K\,(\omega r_0)^\nu$} with \mbox{$\nu\approx 6.0$}; dotted line:~WKB prediction with \mbox{$\nu=4\sqrt{2}\approx 5.66$} from the core-plus-tail action decomposition derived in the main text. The steeper observed exponent reflects the dominance of the throat core region in the WKB action at finite frequencies (see text).}
\label{fig:lowfreq_scaling}
\end{figure}

A log-log plot of $T(\omega)$ for $\ell=1$ over the sub-barrier range \mbox{$0.1\le\omega r_0\le 1.0$} (Fig.~\ref{fig:lowfreq_scaling}) yields an approximately linear trend, consistent with the power-law scaling~\eqref{eq:TWKB_powerlaw}. The effective slope in this range is steeper than the strict asymptotic prediction \mbox{$\nu=2\ell+1=3$}~ \cite{Newton1982}. A log-log fit to the Numerov results over \mbox{$0.003\le\omega r_0\le 0.1$} yields \mbox{$\nu\approx 6.0\pm 0.3$} for \mbox{$\ell=1$}. This steeper exponent reflects contributions from the throat ``core'' region \mbox{($|\sigma|\lesssim r_0$)} to the WKB action at finite frequencies\emdash the strict asymptotic value \mbox{$\nu=2\ell+1$} is expected to be recovered only in the limit \mbox{$\omega r_0\to 0$}~\cite{LiDai2021,Newton1982}. This effective exponent can be understood analytically by decomposing the WKB action \mbox{$S=\int\sqrt{V-\omega^2}\,d\sigma$} into a ``core'' contribution \mbox{$|\sigma|\lesssim r_0$} and a ``tail'' contribution \mbox{$r_0\lesssim|\sigma|\lesssim\sigma_{\mathrm{tp}}$}. The core contributes \mbox{$S_{\mathrm{core}}\approx 2.5$} (approximately independent of $\omega$). The tail, where \mbox{$V\approx\ell(\ell+1)/\sigma^2$}, contributes \mbox{$S_{\mathrm{tail}}\approx 2\sqrt{\ell(\ell+1)}\,\ln(\sigma_{\mathrm{tp}}/r_0)\approx 2\sqrt{\ell(\ell+1)}\,\ln(1/(\omega r_0))$}.

The resulting \mbox{$T\sim\exp(-2S)\sim (\omega r_0)^{4\sqrt{\ell(\ell+1)}}$} predicts an effective exponent \mbox{$\nu=4\sqrt{2}\approx 5.66$} for \mbox{$\ell=1$}, in good agreement with the observed \mbox{$\nu\approx 6.0$}. The strict asymptotic value \mbox{$\nu=2\ell+1=3$}~\cite{Newton1982,LandauLifshitz_QM} is recovered only in the limit \mbox{$\omega r_0\to 0$} where the exact Bessel solutions, rather than WKB, control the transmission. The crossover from the WKB-dominated effective exponent to the strict asymptotic $2\ell+1$ behaviour occurs at frequencies well below our numerical range \mbox{($\omega r_0\lesssim 10^{-3}$)}, and a self-contained Bessel-function matching recovering the \mbox{$\omega^{2\ell+1}$} scaling in that regime is beyond the scope of the present analysis.

At the barrier top \mbox{$\omega=\omega_{\max}^{(\ell=1)}=\sqrt{2}/r_0$}, Numerov integration yields \mbox{$T\approx 0.68$}, noticeably above the value \mbox{$T=1/2$} that would obtain for an inverted-parabola (quadratic) barrier summit~\cite{LandauLifshitz_QM}. The parabolic-top prediction assumes that the wavelength at the summit is small compared to the length scale over which the true potential is well approximated by its quadratic expansion about the maximum. For the EB potential \mbox{$V_\ell(\sigma)=\ell(\ell+1)/(\sigma^2+r_0^2)$}, the parabolic expansion \mbox{$V\approx \ell(\ell+1)/r_0^2-\ell(\ell+1)\sigma^2/r_0^4+\order(\sigma^4)$} is accurate only for \mbox{$|\sigma|\lesssim r_0$}, while the wavelength at the summit is \mbox{$\lambda=2\pi/\omega_{\max}=\pi\sqrt{2}\,r_0\approx 4.4\,r_0$}, comparable to the full width of the barrier core. The wave therefore samples the full non-parabolic barrier shape, including the \mbox{$1/\sigma^2$} tails on both sides, and the parabolic-top result is not expected to apply. As a consistency check, a P\"oschl-Teller barrier \mbox{$V_{\mathrm{PT}}(\sigma)=2\,\mathrm{sech}^2(\sigma/r_0)$} matched to the EB barrier in height and curvature at the summit yields \mbox{$T\approx 0.64$} at the same reduced frequency, confirming that the \mbox{$\sim 0.7$} value is a generic feature of non-parabolic summit shapes rather than a numerical artefact of our Numerov propagation. The numerical unitarity check \mbox{$|R|^2+T=1$} at the summit is satisfied to \mbox{$<10^{-4}$} (Appendix~\ref{app:numerics}).

Replacing plane-wave boundary conditions with exact Bessel asymptotics modifies the extracted coefficients at the few-percent level for \mbox{$\omega r_0\ge 0.01$}, but does not change the qualitative power-law character of the suppression. The sensitivity of \mbox{$T$ to $L_{\max}$} is significant for \mbox{$\omega r_0<0.05$} (where \mbox{$\sigma_{\mathrm{tp}}>30\,r_0$}), but is less than $3\%$ for \mbox{$\omega r_0\ge 0.3$}.

%====================================================================%
\section{Gravitational wave transmission}
\label{sec:GW}

The axial GW effective potential~\eqref{eq:V_GW_EB} differs from the EM potential by the curvature correction
\begin{equation*}
-3r_0^2/(\sigma^2+r_0^2)^2.
\end{equation*}
This term is negative, creating a shallow well superposed on the centrifugal barrier, and reduces the barrier height from \mbox{$\ell(\ell+1)/r_0^2$} to \mbox{$[\ell(\ell+1)-3]/r_0^2$}. For \mbox{$\ell=2$} this gives \mbox{$V_2^{(\mathrm{GW})}(0)=3/r_0^2$}, which is half the EM value \mbox{$V_2^{(\mathrm{EM})}(0)=6/r_0^2$} at the same multipole (Table~\ref{tab:EMvsGW_l2}).

A subtlety arises when comparing across different $\ell$: the GW \mbox{$\ell=2$} barrier, \mbox{$3/r_0^2$}, is actually higher than the EM \mbox{$\ell=1$} barrier, \mbox{$2/r_0^2$}, and the \mbox{$\ell=2$} mode encounters a wider classically forbidden region. The comparison at fixed $\ell$ (Table~\ref{tab:EMvsGW_l2}) more directly illustrates the effect of the curvature correction than the cross-$\ell$ comparison in Table~\ref{tab:TGW}, which conflates the curvature correction with the change in barrier width from different $\ell$.

Table~\ref{tab:TGW} shows the GW transmission coefficients computed by Numerov integration. At \mbox{$\omega r_0=0.5$}, \mbox{$T_{\mathrm{GW}}^{(\ell=2)}\approx 4\times 10^{-5}$} compared with \mbox{$T_{\mathrm{EM}}^{(\ell=1)}\approx 6\times 10^{-4}$}, which shows that the GW mode is more strongly suppressed despite the lower barrier\emdash because the \mbox{$\ell=2$} barrier is wider.

All propagating radiation\emdash EM and GW\emdash is strongly suppressed below its respective barrier-top frequency. The constraint-wave asymmetry is between the static (constraint) and dynamical (wave) sectors, not between specific spin fields.

\begin{table}[t]
\caption{GW (axial, $\ell=2$) transmission on the EB throat ($r_0=1$), compared with EM ($\ell=1$).}
\label{tab:TGW}
\begin{ruledtabular}
\begin{tabular}{cccc}
$\omega r_0$ & $T_{\mathrm{GW}}^{(\ell=2)}$ & $T_{\mathrm{EM}}^{(\ell=1)}$ & $T_{\mathrm{GW}}/T_{\mathrm{EM}}$ \\
\hline
0.1 & $<10^{-15}$ & $2.2\times 10^{-8}$ & $\ll 1$ \\
0.5 & $3.7\times 10^{-5}$ & $5.4\times 10^{-4}$ & $0.069$ \\
1.0 & $3.5\times 10^{-2}$ & $9.5\times 10^{-2}$ & $0.37$ \\
$\sqrt{3}$ & $0.68$ & $0.97$ & $0.70$ \\
2.0 & $0.93$ & $0.99$ & $0.94$ \\
\end{tabular}
\end{ruledtabular}
\end{table}

We have treated only the axial (odd-parity) GW sector. On vacuum backgrounds (Schwarzschild), axial and polar sectors are exactly isospectral~\cite{Chandrasekhar1983}. On the EB background, the polar sector couples to perturbations of the phantom scalar field that sources the geometry~\cite{KimSung2008,Bronnikov2018}, breaking this isospectrality and introducing additional coupled degrees of freedom. However, the centrifugal term \mbox{$\ell(\ell+1)/a^2$} remains the dominant contribution to the polar effective potential for \mbox{$\ell\ge 2$} at the throat. To make this concrete, the polar master equation on an ultrastatic background with a scalar source takes the form~\cite{KimSung2008}
\begin{equation}
\psi'' + \left[\omega^2 - V_{\ell}^{(\mathrm{polar})}\right]\psi = 0,
\end{equation}
where \mbox{$V_{\ell}^{(\mathrm{polar})}$} contains the centrifugal term \mbox{$\ell(\ell+1)/a^2$}, plus a curvature correction and a scalar-field coupling. Near the throat \mbox{($\sigma\approx 0$)}, \mbox{$a\approx r_0$}, and the centrifugal term dominates:
\begin{equation*}
V_{\ell}^{(\mathrm{polar})}(0)\approx \ell(\ell+1)/r_0^2 + \order(1/r_0^2).
\end{equation*}
The scalar coupling modifies the subleading terms, but cannot eliminate the leading \mbox{$\ell(\ell+1)/r_0^2$} barrier. We therefore expect sub-barrier suppression for all propagating GW multipoles to hold in both parity sectors, with the polar sub-barrier transmission $T_{\mathrm{GW,pol}}(\omega)$ differing from the axial value $T_{\mathrm{GW,ax}}(\omega)$ only in the subleading curvature-plus-scalar contributions and therefore at most in the numerical prefactor of the sub-barrier tunnelling action. The qualitative conclusion\emdash strong sub-barrier suppression of all \mbox{$\ell\ge 2$} gravitational wave modes\emdash is unaffected by the choice of parity sector. A full quantitative analysis of the polar sector, including the phantom scalar coupling and the numerical comparison $T_{\mathrm{GW,pol}}/T_{\mathrm{GW,ax}}$ at fixed $\ell$, is left to future work.

\begin{table}[t]
\caption{Like-for-like comparison: EM vs.\ GW transmission at $\ell=2$ on the EB throat ($r_0=1$).
The GW barrier is lower by a factor of~2 due to the curvature correction.}
\label{tab:EMvsGW_l2}
\begin{ruledtabular}
\begin{tabular}{cccc}
$\omega r_0$ & $T_{\mathrm{EM}}^{(\ell=2)}$ & $T_{\mathrm{GW}}^{(\ell=2)}$ & $T_{\mathrm{GW}}/T_{\mathrm{EM}}$ \\
\hline
0.5 & $4.9\times 10^{-9}$ & $9.4\times 10^{-8}$ & 19 \\
1.0 & $9.9\times 10^{-6}$ & $2.9\times 10^{-4}$ & 29 \\
1.5 & $1.4\times 10^{-3}$ & $8.0\times 10^{-2}$ & 56 \\
2.0 & $6.8\times 10^{-2}$ & $0.95$ & 14 \\
$\sqrt{6}$ & $0.61$ & $1.00$ & 1.6 \\
\end{tabular}
\end{ruledtabular}
\end{table}

We compare EM and GW transmission at the same \mbox{$\ell=2$} in Table~\ref{tab:EMvsGW_l2} to highlight the effect of the curvature correction, since this comparison fixes $\ell$ and varies only the field type. The GW barrier height, \mbox{$3/r_0^2$}, is half the EM value, \mbox{$6/r_0^2$}, due to the curvature correction, producing GW transmission \mbox{$\sim\!20$--$55\times$} larger at sub-barrier frequencies.

%====================================================================%
\section{Static gravitational monopole}
\label{sec:monopole}

\subsection{Exact monopole solution on the Ellis-Bronnikov throat}

On the EB throat, Eq.~\eqref{eq:monopole_ultrastatic} becomes \mbox{$((\sigma^2+r_0^2)\Phi')'=0$}, with solution
\begin{equation}
\Phi(\sigma) = \frac{C}{r_0}\arctan\!\left(\frac{\sigma}{r_0}\right)+\Phi_0\,.
\label{eq:monopole_sol}
\end{equation}
The conserved flux \mbox{$\mathcal{F}=a^2\Phi'=C$} is constant throughout, including through the throat. The total potential difference is \mbox{$\Delta\Phi=C\pi/r_0$}.

The identification of $C$ with the enclosed mass follows from Gauss's law.
The outward gravitational flux through a $2$-sphere at position~$\sigma$ is
\begin{equation}
\oint_{S^2}\!\nabla\Phi\cdot dA = 4\pi a(\sigma)^2\Phi'(\sigma) = 4\pi C,
\label{eq:gauss_integral}
\end{equation}
which is independent of~$\sigma$, confirming flux conservation through the throat. For a point mass~$M$ enclosed at $\sigma>0$, Gauss's law gives \mbox{$C=-M$} (attractive convention).

For \mbox{$d\gg r_0$}, Newtonian matching at the source determines this constant, and the potential on the far side of the throat is
\begin{equation}
\Phi(-d) \approx -\frac{M}{d}\!\left(1-\frac{\pi r_0}{2d}+\cdots\right),
\label{eq:newtonian_matching}
\end{equation}
with corrections of order \mbox{$r_0/d$}. At leading order, the gravitational potential on the far side of the throat is indistinguishable from that of a mass $M$ in flat space\emdash the throat introduces only a subleading correction proportional to the ratio \mbox{$r_0/d$} of the throat radius to the source distance.

\subsection{Higher gravitational multipoles}
\label{sec:higher_multipoles}

The monopole, \mbox{$\ell=0$}, transmits because it encodes the total enclosed mass, whose gravitational flux is conserved through any closed surface\emdash including one that threads the throat. This conservation is reflected in the absence of a centrifugal term in the \mbox{$\ell=0$} Poisson equation. We now consider higher multipoles \mbox{($\ell\ge 1$)} of the gravitational field.

The \mbox{$\ell\ge 1$} components of the Poisson equation on the spatial metric \mbox{$d\sigma^2+a^2 d\Omega^2$} satisfy, in the source-free region,
\begin{equation}
\frac{1}{a^2}\frac{d}{d\sigma}\!\left(a^2\frac{d\Phi_\ell}{d\sigma}\right) - \frac{\ell(\ell+1)}{a^2}\Phi_\ell = 0.
\label{eq:higher_multipoles}
\end{equation}
This can be rewritten in Schr\"odinger form by substituting \mbox{$\psi_\ell=a\,\Phi_\ell$}:
\begin{equation}
\psi_\ell'' - \left[\frac{\ell(\ell+1)}{a^2} + \frac{a''}{a}\right]\psi_\ell = 0.
\label{eq:higher_schrodinger}
\end{equation}
For \mbox{$\ell\ge 1$}, the centrifugal term \mbox{$\ell(\ell+1)/a^2$} dominates near the throat and creates a barrier\emdash the same type of barrier that suppresses EM and GW modes. These modes are not propagating waves (they are static solutions of an elliptic equation), but they still decay through the barrier region, qualitatively like evanescent EM modes.

The gravitational dipole, \mbox{$\ell=1$}, corresponds to the tidal field, whereas the quadrupole, \mbox{$\ell=2$}, corresponds to the next order of gravitational structure. Both decay through the throat. Only the monopole\emdash the total enclosed mass\emdash is transmitted without sub-barrier suppression.

To quantify the decay, we note that Eq.~\eqref{eq:higher_schrodinger} at \mbox{$\omega^2=0$} has the form of a zero-energy scattering problem with effective potential
\begin{equation*}
U_{\mathrm{eff}}=\ell(\ell+1)/a^2+a''/a.
\end{equation*}
Near the throat \mbox{($|\sigma|\lesssim r_0$)}, the centrifugal term
\begin{equation*}
\ell(\ell+1)/a^2\approx\ell(\ell+1)/r_0^2
\end{equation*}
dominates over the curvature correction 
\begin{equation*}
a''/a=r_0^2/(\sigma^2+r_0^2)^2\le 1/r_0^2,
\end{equation*}
so the solution decays through the throat region. At large \mbox{$|\sigma|$}, the two independent solutions behave as \mbox{$\sigma^\ell$} (growing) and \mbox{$\sigma^{-(\ell+1)}$} (decaying).

A source at \mbox{$\sigma=d\gg r_0$} produces a field \mbox{$\Phi_\ell\propto d^{-(\ell+1)}$}\emdash the component transmitted to \mbox{$\sigma=-d$} involves coupling through the throat and scales as \mbox{$(r_0/d)^{2\ell+1}$} relative to the source-side amplitude. This scaling follows from standard multipole matching across a constriction~\cite{Visser1995,Jackson1998}. On the source side, the field at the throat \mbox{($\sigma\sim 0$)} is \mbox{$\Phi_\ell\sim d^{-(\ell+1)} r_0^\ell$} which is the \mbox{$\ell$-th} multipole evaluated at distance~$d$ and projected onto the throat scale~$r_0$. On the far side, the growing solution \mbox{$\sigma^\ell$} couples with amplitude \mbox{$\sim r_0^\ell/d^{\ell+1}$}, giving
\begin{equation*}
\Phi_\ell(\text{far})/\Phi_\ell(\text{source})\sim (r_0/d)^{2\ell+1}.
\end{equation*}
For \mbox{$\ell=1$} at \mbox{$d=10\,r_0$}, this gives a suppression factor \mbox{$\sim(r_0/d)^3=10^{-3}$}, and for \mbox{$\ell=2$}, \mbox{$\sim(r_0/d)^5=10^{-5}$}. These are polynomial, rather than exponential, suppressions, but they are much stronger than the monopole \mbox{($\ell=0$)} correction of order \mbox{$r_0/d$}.

We have verified this scaling numerically by integrating Eq.~\eqref{eq:higher_multipoles} from the throat outward with even initial conditions (\mbox{$\Phi(0)=1$}, \mbox{$\Phi'(0)=0$}). At large \mbox{$|\sigma|$}, the solution grows as \mbox{$\Phi\sim C_\ell\,\sigma^\ell$}\emdash we extract \mbox{$C_\ell$} and verify the exponent by fitting \mbox{$\ln\Phi$} vs \mbox{$\ln\sigma$} at \mbox{$\sigma/r_0=10$--$300$}. Table~\ref{tab:higher_multipoles} summarises the results: the fitted growth exponent matches $\ell$ to better than $0.01\%$ for all cases, and the connection coefficients $C_\ell$ are \mbox{$\order(1)$}.
The $C_\ell$ values reported in Table~\ref{tab:higher_multipoles} are specific to the Ellis-Bronnikov profile, but the scaling~\eqref{eq:higher_multipoles} is generic: on any throat background with areal radius minimum $r_0$ and finite lapse at the throat, integrating Eq.~\eqref{eq:higher_multipoles} outward with even initial conditions produces a growing solution $C_\ell\sigma^\ell$ at large $|\sigma|$ with an $\order(1)$ coefficient depending on the detailed shape of $a(\sigma)$ through the near-throat region.
For the sub-topological profiles considered in this paper, we find that $C_\ell$ varies by at most a factor of a few across the parametric family and the reflected-Schwarzschild wormhole; the qualitative scaling \mbox{$(r_0/d)^{2\ell+1}$} is therefore robust, with only the numerical prefactor dependent on throat details.
The \mbox{$(r_0/d)^{2\ell+1}$} suppression then follows from combining the source-side decay, \mbox{$d^{-(\ell+1)}$}, with the throat-scale connection coefficient \mbox{$C_\ell$} and the far-side growing mode \mbox{($d^\ell$)}.

\begin{table}[h]
\caption{Connection coefficients $C_\ell$ and fitted growth exponents for the even solution of the static higher-multipole equation~\eqref{eq:higher_multipoles} on the EB throat.}
\label{tab:higher_multipoles}
\begin{ruledtabular}
\begin{tabular}{cccc}
$\ell$ & $C_\ell$ & Fitted exponent & Expected \\
\hline
1 & $\pi/2 \approx 1.5708$ & $1.0000$ & $1$ \\
2 & $3.0000$ & $2.0000$ & $2$ \\
3 & $5.8905$ & $3.0000$ & $3$ \\
\end{tabular}
\end{ruledtabular}
\end{table}

This sharpens the characterisation of the asymmetry. It is not that electromagnetic fields are suppressed while gravitational fields are not, but all fields with \mbox{$\ell\ge 1$}\emdash electromagnetic, gravitational wave, and higher gravitational multipoles alike\emdash are suppressed by the centrifugal barrier at the throat. Only the \mbox{$\ell=0$} gravitational monopole, the conserved charge of the gravitational field, transmits smoothly. The asymmetry is between the constraint (monopole, Gauss's law) and all radiative and tidal degrees of freedom.

\subsection{Physical interpretation}

The monopole solution~\eqref{eq:monopole_sol} solves an elliptic equation, not a hyperbolic one. There is no ``incident wave,'' no ``reflection,'' and no ``transmission coefficient'' in the wave-scattering sense. The relevant quantity is the conserved flux \mbox{$\mathcal{F}=a^2\Phi'$}, which is the content of Gauss's law on the curved background\emdash the gravitational flux through any closed surface surrounding the source is determined by the enclosed mass, independent of the intervening geometry.

This conservation law is exact and non-perturbative. It does not depend on the specific profile $a(\sigma)$, on whether the geometry satisfies the energy conditions, or on the source of the spacetime curvature. On non-vacuum backgrounds such as EB (sourced by a phantom scalar), the \mbox{$\ell=0$} gravitational perturbation $\Phi$ is gauge-invariant at linear order: it represents the physical Newtonian potential (the fractional lapse perturbation \mbox{$\delta N/N$}), and the background scalar field perturbation decouples from the \mbox{$\ell=0$} sector of the lapse equation because the Hamiltonian constraint absorbs the background stress-energy terms (see Appendix~\ref{app:monopole}). On non-ultrastatic backgrounds such as the DS construction, \mbox{$\Phi=\delta N/N_0$} retains its interpretation as the fractional lapse perturbation\emdash a distant observer measures it through the redshifted Newtonian potential $N_0\Phi$, which encodes the physical gravitational acceleration. It is a consequence of the structure of the Laplace equation on a manifold with a minimal-area surface connecting two asymptotic regions.

For quantitative comparison with the EM sector, we define a ``monopole ratio'' as the ratio of the actual gravitational potential at distance~$d$ on the far side of the throat to the Newtonian value that would obtain in flat space at the same distance:
\begin{equation}
\mathcal{T}_{\mathrm{mono}}(d) \equiv \frac{\Phi_{\mathrm{throat}}(-d)}{\Phi_{\mathrm{flat}}(d)} = \frac{(M/r_0)[\pi/2-\arctan(d/r_0)]}{M/d}.
\label{eq:T_mono}
\end{equation}
For \mbox{$d\gg r_0$}, Taylor-expanding the arctan gives \mbox{$\mathcal{T}_{\mathrm{mono}}\to 1-\order(r_0/d)$}\emdash the potential on the far side approaches the flat-space Newtonian value with corrections of order~\mbox{$r_0/d$}. At \mbox{$d=10\,r_0$}, \mbox{$\mathcal{T}_{\mathrm{mono}}\approx 0.997$}, and \mbox{$\approx 1.000$} at \mbox{$d=100\,r_0$}.
This polynomial attenuation is the defining feature of the constraint-wave asymmetry\emdash the growing ratio \mbox{$\mathcal{T}_{\mathrm{mono}}/T_{\mathrm{EM}}^{(\ell=1)}$} is shown in the lower panel of Fig.~\ref{fig:scattering}.

%====================================================================%
\section{Universality across throat shapes}
\label{sec:universality}

\subsection{Parametric family}

For the parametric family~\eqref{eq:parametric}, the EM effective potential is
\begin{equation}
V_\ell^{(n)}(\sigma) = \frac{\ell(\ell+1)}{r_0^2}\!\left(1+\frac{\sigma^2}{r_0^2}\right)^{-1/n}\!\!.
\label{eq:V_parametric}
\end{equation}
The barrier height \mbox{$\ell(\ell+1)/r_0^2$} is $n$-independent\emdash the width increases with~$n$. The asymptotic tail also changes, and at large $|\sigma|$
\begin{equation*}
a_n\sim r_0(\sigma/r_0)^{1/n},
\end{equation*}
so
\begin{equation*}
V_\ell^{(n)}\sim\ell(\ell+1)\,r_0^{2/n-2}\,|\sigma|^{-2/n}.
\end{equation*}

For $n=1$ (EB), the tail is \mbox{$\propto 1/\sigma^2$}, producing power-law suppression \mbox{($T\sim\omega^\nu$)}. For \mbox{$n\ge 2$}, the tail decays more slowly \mbox{($\propto 1/|\sigma|^{2/n}$)}, causing the WKB turning points to recede as \mbox{$\sigma_{\mathrm{tp}}\sim\omega^{-n}$}, and the action to grow as \mbox{$\omega^{-(n-1)}$}. The resulting transmission is stretched-exponential:
\begin{equation*}
T\sim\exp(-\mathrm{const}/\omega^{n-1}).
\end{equation*}
This explains the dramatic suppression seen in Table~\ref{tab:parametric} for \mbox{$n=2$} and \mbox{$n=4$}, which is far stronger than any power law. The monopole equation \mbox{$(a_n^2\Phi')'=0$} has the smooth solution
\begin{equation*}
\Phi=C\int_0^\sigma d\sigma'/a_n^2+\Phi_0, \quad \forall n.
\end{equation*}

Table~\ref{tab:parametric} shows the EM tunnelling amplitude at several frequencies for \mbox{$n=1,2,4$}. (For \mbox{$n=1$}, which is asymptotically flat, this coincides with the standard scattering transmission coefficient.) The suppression increases dramatically with~$n$\emdash at \mbox{$\omega r_0=0.1$}, $T$ drops from \mbox{$\sim\!10^{-8}$} \mbox{($n=1$)} to below \mbox{$10^{-40}$} \mbox{($n=4$)}, reflecting the transition from a narrow \mbox{($n=1$)} to a broad \mbox{($n=4$)} barrier. The monopole flux is conserved identically.

\begin{table}[h]
\caption{EM tunnelling amplitude ($\ell=1$) for the parametric throat family at selected frequencies. Barrier height is $n$-independent; width increases with~$n$.}
\label{tab:parametric}
\begin{ruledtabular}
\begin{tabular}{ccccc}
$n$ & $\omega r_0=0.1$ & $\omega r_0=0.3$ & $\omega r_0=0.5$ & $\omega r_0=1.0$ \\
\hline
1 & $2.1\times 10^{-8}$ & $1.9\times 10^{-5}$ & $5.4\times 10^{-4}$ & $8.8\times 10^{-2}$ \\
2 & $5.0\times 10^{-36}$ & $5.1\times 10^{-16}$ & $1.1\times 10^{-8}$ & $1.2\times 10^{-2}$ \\
4 & $<10^{-40}$ & $2.1\times 10^{-54}$ & $3.8\times 10^{-40}$ & $1.5\times 10^{-4}$ \\
\end{tabular}
\end{ruledtabular}
\end{table}

As a complementary benchmark, we consider a finite-length cylindrical neck:
\begin{equation*}
a(\sigma) = 
\begin{cases}
r_0,      & |\sigma|\le L/2 \\
|\sigma|, & |\sigma|>L/2
\end{cases}
\end{equation*}
creating a compact-support barrier. The WKB prediction is \mbox{$T\sim\exp(-2\sqrt{\ell(\ell+1)}\,L/r_0)$}. Numerov integration confirms this\emdash for \mbox{$\ell=1$} and \mbox{$L=10\,r_0$}, \mbox{$T=3.2\times 10^{-13}$}, compared with the WKB prediction 
\mbox{$\exp(-2\sqrt{2}\times 10)=6.4\times 10^{-13}$}. The agreement validates the numerical method and illustrates the transition from power-law (long-range tail) to exponential (compact support) suppression.

For \mbox{$n\ge 2$}, the profile \mbox{$a_n(\sigma)\sim|\sigma|^{1/n}$} at large $|\sigma|$ is not asymptotically flat \mbox{($a/|\sigma|\to 0$} rather than~$1$). Plane-wave scattering boundary conditions are therefore not strictly justified for \mbox{$n\ge 2$}\emdash the values in Table~\ref{tab:parametric} should be interpreted as tunnelling amplitudes computed on a finite domain rather than standard transmission coefficients in the asymptotically flat sense.
A rigorous treatment would replace plane waves at the domain boundary with exact solutions matched to the long-range form of the potential (e.g., matched asymptotics against the Bessel-like solutions governed by the $\sim\!1/a^2$ centrifugal tail), which would shift the extracted amplitudes at the percent level but leave the qualitative stretched-exponential dependence on $n$ unchanged.
We have not performed this matching here; the systematic error this introduces is bounded below by the sensitivity of our finite-domain results to the boundary location $L_{\max}$, which for the sub-barrier frequencies tabulated is \mbox{$\lesssim 5\%$}.
Our universality claims regarding the qualitative \mbox{$\ell=0$} versus \mbox{$\ell\ge 1$} asymmetry rest on the asymptotically flat cases (EB and the reflected-Schwarzschild wormhole), not on the parametric family at \mbox{$n\ge 2$}.

\subsection{Transmission on the reflected-Schwarzschild background}
\label{sec:DS_wormhole_analysis}

The DS wormhole~\cite{DamourSolodukhin2007} is constructed by taking the Schwarzschild exterior ($a>r_0$) and reflecting it across $a=r_0$ to create a second asymptotic region, replacing the would-be horizon with a throat (Fig.~\ref{fig:profiles}, right panel). In our proper-distance coordinate, the metric on each side is~\eqref{eq:DS} with \mbox{$a\ge r_0$}\emdash the two copies are joined at \mbox{$\sigma=0$} ($a=r_0$), giving \mbox{$\sigma\in(-\infty,+\infty)$}. The proper distance from the throat to areal radius~$a$ is
\begin{align*}
\sigma&=\int_{r_0}^a da'/\sqrt{1-r_0/a'}\\
      &=\sqrt{a(a-r_0)}+r_0\ln[(\sqrt{a}+\sqrt{a-r_0})/\sqrt{r_0}].
\end{align*}
The redshift factor is \mbox{$e^{2\alpha}=1-r_0/a(\sigma)$}, which vanishes at the throat.

\begin{figure*}[t]
\centering
\includegraphics[width=\textwidth]{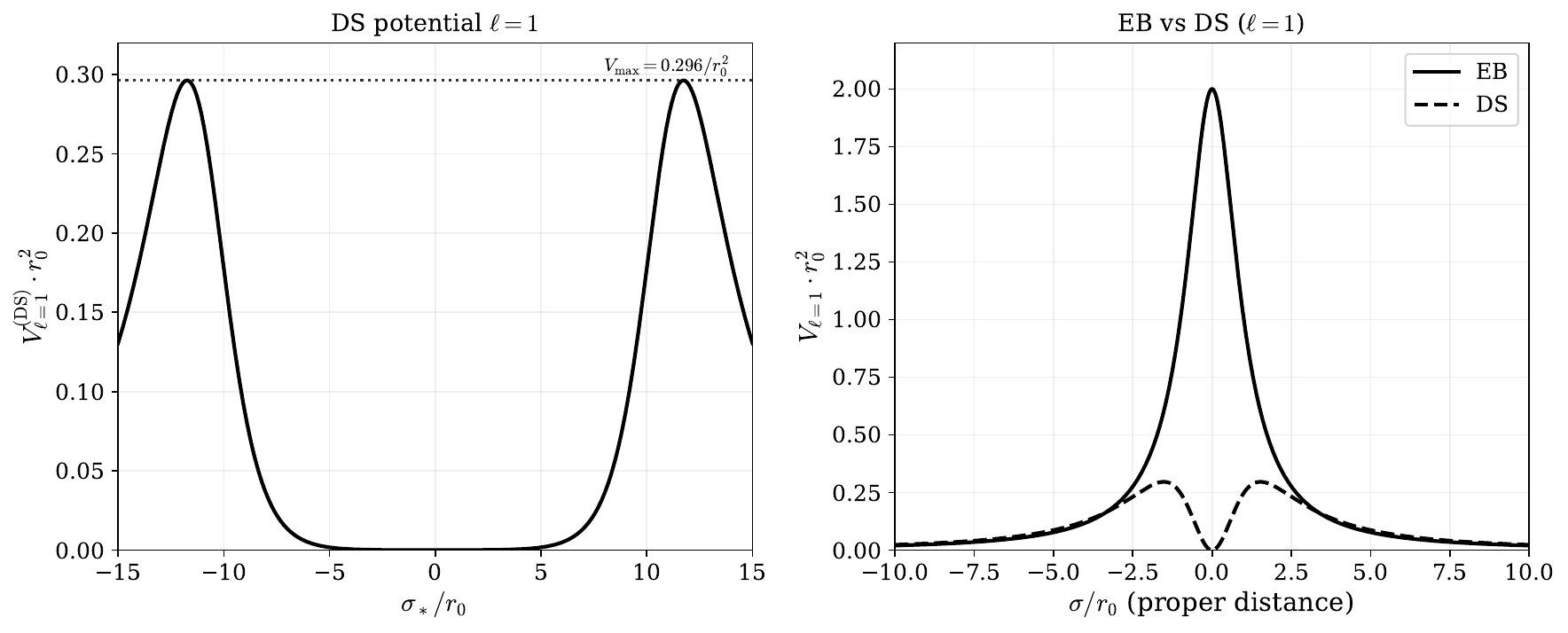}
\caption{Left panel:~EM effective potential ($\ell=1$) for the DS wormhole in the tortoise coordinate, showing the double-peak structure with barrier top at $V_{\max}=8/(27r_0^2)\approx 0.296/r_0^2$ (dashed line).
Right panel:~Comparison of EB (single peak, taller) and DS (double peak, shorter) potentials in proper-distance coordinates.}
\label{fig:DS}
\end{figure*}

The EM effective potential in the tortoise coordinate $\sigma_*$ (with \mbox{$d\sigma_*=e^{-\alpha}d\sigma=(1-r_0/a)^{-1/2}\,d\sigma$}) is
\begin{equation}
V_\ell^{(\mathrm{DS})} = \left(1-\frac{r_0}{a}\right)\frac{\ell(\ell+1)}{a^2}\,.
\label{eq:V_DS}
\end{equation}
This vanishes at the throat \mbox{($a=r_0$)}, rises to a maximum \mbox{$V_\ell^{\max}=4\ell(\ell+1)/(27r_0^2)$} at the photon sphere \mbox{$a=3r_0/2$}, and decays as \mbox{$\ell(\ell+1)/a^2$} at large~$a$ (Fig.~\ref{fig:DS}, left panel). The DS barrier is a double-peaked structure (one peak on each side of the throat), unlike the single-peaked EB barrier (Fig.~\ref{fig:DS}, right panel).

We solve the scattering problem in the tortoise coordinate using Numerov integration, with boundary conditions

\begin{equation*}
\Psi\to \begin{cases}
e^{-i\omega\sigma_*}+R\,e^{+i\omega\sigma_*}, & \sigma_*\to+\infty \\
A\,e^{-i\omega\sigma_*},                      & \sigma_*\to-\infty
\end{cases}
\end{equation*}

\begin{table}[b]
\caption{EM transmission \mbox{($\ell=1$)} on the Damour-Solodukhin wormhole \mbox{($r_0=1$)}. The barrier-top frequency is \mbox{$\omega_{\max}\approx 0.544/r_0$}.}
\label{tab:DS}
\begin{ruledtabular}
\begin{tabular}{ccc}
$\omega r_0$ & $T_{\mathrm{EM}}^{(\ell=1)}$ (DS) & $T_{\mathrm{EM}}^{(\ell=1)}$ (EB) \\
\hline
0.1 & $\sim 10^{-6}$ & $2.2\times 10^{-8}$ \\
0.3 & $\sim 10^{-2}$ & $1.8\times 10^{-5}$ \\
0.5 & $0.6$ & $5.8\times 10^{-4}$ \\
1.0 & $0.99$ & $9.5\times 10^{-2}$ \\
\end{tabular}
\end{ruledtabular}
\end{table}

The EM transmission coefficients for \mbox{$\ell=1$} are shown in Table~\ref{tab:DS}. The qualitative behaviour matches the EB case: strong sub-barrier suppression, with \mbox{$T\to 1$} above the barrier top. The barrier-top frequency is
\begin{equation*}
\omega_{\max}^{(\ell=1)}=\sqrt{V_1^{\max}}=2\sqrt{2}/(3\sqrt{3}\,r_0)\approx 0.544/r_0,
\end{equation*}
lower than the EB cutoff $\sqrt{2}/r_0$ due to the redshift suppression. The DS barrier is lower and narrower than the EB barrier (due to the redshift factor suppressing the potential), so the sub-barrier suppression is weaker.

For the GW axial sector on the reflected-Schwarzschild background, the Regge-Wheeler potential takes the form
\begin{equation*}
V_\ell^{(\mathrm{GW})}\big|_{\mathrm{DS}}=(1-r_0/a)[\ell(\ell+1)/a^2-3r_0/a^3],
\end{equation*}
which has the same double-peak structure as the EM potential. For \mbox{$\ell=2$} the barrier maximum is at \mbox{$a\approx 1.640\,r_0$} (shifted from the Schwarzschild photon sphere at \mbox{$3r_0/2$} by the curvature correction) with value \mbox{$V_{\max}\approx 0.605/r_0^2$}, giving the barrier-top frequency \mbox{$\omega_{\max}^{(\mathrm{GW},\ell=2)}\approx 0.778/r_0$}. Numerov integration of the GW master equation in the tortoise coordinate yields the transmission coefficients in Table~\ref{tab:DS_GW}. The qualitative behaviour matches the EM case: strong sub-barrier suppression, rising to \mbox{$T\to 1$} above the barrier top. At \mbox{$\omega r_0=0.5$}, for example, \mbox{$T_{\mathrm{GW}}^{(\ell=2)}\approx 1.6\times 10^{-3}$}, many orders of magnitude smaller than the corresponding EM value \mbox{$T_{\mathrm{EM}}^{(\ell=1)}\approx 0.6$} because the GW \mbox{$\ell=2$} mode lies deeper below its barrier top than the EM \mbox{$\ell=1$} mode does below its own. The qualitative sub-barrier suppression\emdash and therefore the constraint-wave asymmetry of the static monopole against all propagating multipoles\emdash holds on the DS background for both EM and GW sectors.

\begin{table}[h]
\caption{Axial GW transmission \mbox{($\ell=2$)} on the Damour-Solodukhin wormhole at \mbox{$\lambda\to 0$} \mbox{($r_0=1$)}. The barrier-top frequency is \mbox{$\omega_{\max}^{(\mathrm{GW},\ell=2)}\approx 0.778/r_0$}. Numerov integration in the tortoise coordinate, \mbox{$N=4\times 10^4$}, \mbox{$L_{\max}=80\,r_0$}; unitarity satisfied to \mbox{$<10^{-6}$} at every tabulated frequency.}
\label{tab:DS_GW}
\begin{ruledtabular}
\begin{tabular}{cc}
$\omega r_0$ & $T_{\mathrm{GW}}^{(\ell=2)}$ (DS) \\
\hline
$0.1$          & $5.8\times 10^{-12}$ \\
$0.3$          & $1.3\times 10^{-6}$ \\
$0.5$          & $1.6\times 10^{-3}$ \\
$0.778$        & $0.53$ \\
$1.0$          & $0.77$ \\
$1.5$          & $0.997$ \\
\end{tabular}
\end{ruledtabular}
\end{table}

The double-peak structure of the DS barrier could, in principle, support resonant tunnelling at specific frequencies corresponding to quasi-bound states trapped between the two peaks~\cite{CardosoPani2019}. Such resonances would appear as narrow transmission peaks at or above the barrier-top frequency. A coarse frequency-domain scan over \mbox{$0.1\le\omega r_0\le 0.8$} (200 frequency points) for \mbox{$\ell=1$} reveals no narrow resonance peaks in the deep sub-barrier regime\emdash the transmission increases monotonically with frequency. This is consistent with the expectation that resonant tunnelling between the two peaks requires at least one peak to be classically allowed \mbox{($\omega\gtrsim\omega_{\max}$)}, and a frequency-domain scan at fixed $\omega$ does not probe the transient ringdown-and-echo structure that would appear in a full time-domain evolution.
Resonances and echoes are expected near and above \mbox{$\omega_{\max}$}~\cite{CardosoPani2019,BuenoEtAl2018,MondalSadhukhan2025}. Explicit time-domain analyses of echoes in the Maldacena-Milekhin-Popov traversable wormhole~\cite{MaldacenaMilekhinPopov2023,MondalSadhukhan2025} show that higher-$\ell$ modes produce stronger echoes, consistent with the $\ell$-dependent barrier heights identified here. A full time-domain evolution of wavepacket scattering on the DS background\emdash which would directly reveal any echo or trapped-mode structure missed by our frequency-domain scan\emdash and the quantitative characterisation of resonance widths, spacing, and $\ell$-dependence are left to future work.

The monopole equation gives \mbox{$a^2(1-r_0/a)\Phi'=C$}, and therefore \mbox{$\Phi'=C/[a^2(1-r_0/a)]$}. Near the throat, \mbox{$a-r_0\propto\sigma^2$}, so \mbox{$\Phi'\propto 1/\sigma^2$}, which is not integrable across \mbox{$\sigma=0$}\emdash the monopole potential $\Phi$ diverges at the throat. This is a consequence of the infinite redshift \mbox{($e^{2\alpha}\to 0$)} at the throat in the \mbox{$\lambda=0$} limit.

For the DS family with \mbox{$\lambda>0$}, the lapse at the throat is \mbox{$e^{2\alpha}=\lambda^2(1+\lambda^2)^{-1}$}, and the monopole gradient \mbox{$\Phi'(0)=\mathcal{F}/(r_0^2\,\lambda^2)$} is finite for any \mbox{$\lambda>0$}\emdash the singularity is absent and the monopole transmits smoothly, just as in the ultrastatic case. As \mbox{$\lambda\to 0$}, \mbox{$\Phi'(0)\to\infty$} and the potential develops a \mbox{$1/|\sigma|$} divergence at the throat, so the smooth-transmission interpretation breaks down. The \mbox{$\lambda\to 0$} limit is therefore non-traversable for the monopole as well as for waves, and falls outside the scope of our universality claim.

The EM sector, by contrast, still encounters a barrier for all \mbox{$\lambda\ge 0$}. The constraint-wave asymmetry therefore persists for non-ultrastatic throats provided the lapse remains nonzero at the throat \mbox{($e^{2\alpha}(\sigma=0)>0$)}, which is satisfied by the standard DS construction with \mbox{$\lambda>0$}.

The transmission coefficients computed at $\lambda=0$ differ from the $\lambda>0$ case only by $\order(\lambda^2)$ corrections in the barrier height, since the centrifugal term and the photon-sphere location are unchanged to leading order in $\lambda$. The qualitative sub-barrier behaviour is therefore unaffected, and the values in Table~\ref{tab:DS} remain representative of the physically traversable $\lambda>0$ case for small $\lambda$.

%====================================================================%
\section{Discussion}

\subsection{Origin of the asymmetry}
\label{sec:asymmetry}

The results of Sections~\ref{sec:EM}--\ref{sec:universality} establish a universal asymmetry in field transmission through throat spacetimes. The origin is the multipole decomposition of the field equations on a spherically symmetric background:

\begin{enumerate}[leftmargin=2em,label=(\roman*)]

\item Propagating fields (EM with \mbox{$\ell\ge 1$}, GW with \mbox{$\ell\ge 2$}, and higher static gravitational multipoles with \mbox{$\ell\ge 1$)} all encounter an effective centrifugal barrier \mbox{$V_\ell\propto\ell(\ell+1)/a^2$}, which peaks at any minimal-area surface. Below the barrier, transmission is strongly suppressed by sub-barrier tunnelling (power-law for long-range tails, exponential for compact barriers).

\item The static gravitational monopole \mbox{($\ell=0$)} satisfies the conservation law \mbox{$(a^2\Phi')'=0$} (ultrastatic) or \mbox{$(a^2 e^{2\alpha}\Phi')'=0$} (general, see Appendix~\ref{app:monopole}), which has no centrifugal barrier. The conserved flux $\mathcal{F}$ is the gravitational analogue of Gauss's law, and its conservation is a topological property of the Laplace equation.

\end{enumerate}

The fundamental mathematical distinction is between the \mbox{$\ell=0$} and \mbox{$\ell\ge 1$} sectors of the Laplacian on a manifold with a minimal-area surface. The centrifugal barrier \mbox{$\ell(\ell+1)/a^2$} vanishes identically for \mbox{$\ell=0$}, and is positive for all \mbox{$\ell\ge 1$}. More precisely, for any static, spherically symmetric throat with \mbox{$a'(0)=0$} and \mbox{$a''(0)>0$}, the function \mbox{$\ell(\ell+1)/a(\sigma)^2$} attains a strict maximum at \mbox{$\sigma=0$} for every \mbox{$\ell\ge 1$}, creating a barrier, while the \mbox{$\ell=0$} equation \mbox{$(a^2 e^{2\alpha}\Phi')'=0$} is a first integral with no extremum to overcome.

These two properties\emdash barrier for \mbox{$\ell\ge 1$}, conservation law for \mbox{$\ell=0$}\emdash depend only on the existence of a minimal-area surface and the spherical symmetry of the decomposition, constituting the formal basis of the universality claim. For fixed-background propagation, this is independent of the specific throat profile $a(\sigma)$ and the matter content sourcing the geometry, provided the lapse remains nonzero at the throat \mbox{($e^{2\alpha}>0$)}. Beyond this kinematic statement, dynamical backreaction, coupling to source fields in the polar GW sector, and stability of the throat geometry could modify the quantitative transmission in practice\emdash the universality claim applies to the kinematic structure of the field equations on a given background.

The spherical symmetry assumption underlying the multipole decomposition is important: on rotating (axisymmetric) throat geometries the $(\ell,m)$ modes couple through frame-dragging and the effective potential acquires an $m$-dependence that can shift barrier heights and lift the degeneracy between axial and polar sectors. A rotating-throat analysis is beyond the scope of the present paper, but the $\ell=0$ monopole conservation law (which invokes no $m$-mixing) is expected to survive the extension to axisymmetric backgrounds whenever the lapse remains bounded away from zero at the throat.

This asymmetry has a close analogy in condensed matter physics. A metallic waveguide with a narrow constriction exhibits the same phenomenology. The DC electrical resistance (the \mbox{$\ell=0$} analogue, governed by Laplace's equation) depends only polynomially on the constriction geometry, while AC signals above a frequency-dependent cutoff propagate freely, and below it are evanescent. The throat in curved spacetime plays the role of the constriction, and the centrifugal barrier plays the role of the waveguide cutoff\emdash but without material walls.

\subsection{Relation to existing work}

The effective potentials~\eqref{eq:V_EM_general} and their applications to wormhole scattering are well-studied. Kim and Kim~\cite{KimSung2008} computed EM and scalar scattering coefficients on the EB throat, obtaining quasinormal mode frequencies and absorption cross sections. Konoplya and collaborators~\cite{Konoplya2005,KonoplyaZhidenko2010} provided a comprehensive analysis of quasinormal modes and scattering for a range of wormhole geometries. Churilova, Konoplya, Stuchlík, and Zhidenko~\cite{Churilova2021} extended this analysis to wormholes supported by physical matter in Einstein-Maxwell-Dirac theory and in Randall-Sundrum brane-world models, including the regime where the quasinormal ringing mimics near-extremal Reissner-Nordström black holes. Aneesh, Bose, and Kar~\cite{Aneesh2018} computed scalar quasinormal modes for a two-parameter family of Lorentzian wormholes in scalar-tensor gravity, identifying breathing-mode signatures accessible to gravitational-wave observations. Cardoso and Pani~\cite{CardosoPani2019} reviewed the phenomenology of compact objects including wormholes, with emphasis on GW echoes. Systematic treatments of long-range scattering boundary conditions have been developed by Li and Dai~\cite{LiDai2021}.

Recent studies have extended wormhole perturbation theory to quantum-corrected geometries~\cite{Konoplya2022qc}, and to families interpolating between black holes and wormholes~\cite{Bronnikov2022}, and and have developed detailed echo analyses for traversable wormholes~\cite{BuenoEtAl2018}. Closely related techniques have been applied to braneworld black holes with partially reflecting near-horizon regions, where the reflection-induced echo structure parallels that of a wormhole throat~\cite{Dey2020}. For the Damour-Solodukhin background in particular, Qian \textit{et al.}~\cite{Qian2024} have analysed the interplay of late-time tails and echoes using a Green's function approach, and subsequent work~\cite{Qian2025} has examined the relation between reflectionless modes and echo modes for symmetric and asymmetric DS wormholes, finding that reflectionless modes provide a complementary characterisation of the same echo phenomenology. The sub-barrier regime addressed here complements this literature by focusing on the strong-suppression limit rather than quasinormal spectra or echo timing.

The novelty of the work presented here is threefold:
\begin{enumerate}[leftmargin=2em,label=(\roman*)]

\item the systematic comparison across EM, GW, and static monopole sectors on the same background, particularly in the sub-barrier regime that is typically not the focus of quasinormal mode studies;

\item the demonstration of universality across a parametric family of throat profiles and non-ultrastatic geometries; and

\item the identification of the asymmetry as structural, arising from the \mbox{$\ell=0$} versus \mbox{$\ell\ge 1$} decomposition of the Laplacian, and the observation that higher gravitational multipoles \mbox{($\ell\ge 1$)} are suppressed just like EM modes, leaving only the monopole as the unsuppressed channel.

\end{enumerate}

The ``monopole ratio'' \mbox{$\mathcal{T}_{\mathrm{mono}}$} (Eq.~\eqref{eq:T_mono}) is conceptually distinct from the wave-scattering transmission coefficient $T(\omega)$\emdash the former characterises a static solution of an elliptic equation, and the latter a dynamical scattering problem for a hyperbolic equation. The monopole is not ``transmitted through'' the throat in the causal sense of a signal propagating from one side to the other\emdash the Poisson equation on the full spatial manifold has a solution in which the potential is nonzero on both sides.

\subsection{Multi-messenger implications}

The constraint-wave asymmetry has immediate implications for multi-messenger observations in the presence of throat geometries. If a source on the far side of a throat emits both electromagnetic and gravitational wave radiation simultaneously (e.g., a binary neutron star merger), the EM signal would be strongly suppressed for frequencies below \mbox{$\omega_{\max}=\sqrt{2}/r_0$}, while the GW signal would be suppressed for frequencies below \mbox{$\omega_{\max}^{(\mathrm{GW},\ell=2)}=\sqrt{3}/r_0$}. The source would appear as a ``GW-loud, EM-quiet'' transient\emdash or, for $r_0$ smaller than the GW wavelength, as ``gravity-only.''

Current LIGO-Virgo-KAGRA observations operate at \mbox{$f\sim 10$--$10^3$}~Hz. GW suppression of the lowest GW multipole (\mbox{$\ell=2$}, barrier top \mbox{$\omega_{\max}=\sqrt{3}/r_0$}) in the LIGO band requires \mbox{$r_0\gtrsim\sqrt{3}\,c/(2\pi f)$}: at \mbox{$f=100$}~Hz, \mbox{$r_0\gtrsim 800$}~km, and \mbox{$r_0\gtrsim 80$}~km at \mbox{$f=1000$}~Hz. The future space-based detector LISA, targeting \mbox{$f\sim 10^{-4}$--$1$}~Hz, would require \mbox{$r_0\gtrsim 8\times 10^4$}~km (at 1~Hz) to \mbox{$8\times 10^8$}~km (at \mbox{$10^{-4}$}~Hz). For microscopic throats \mbox{($r_0\ll 1$}~m), all astrophysically observed GWs would propagate freely.

For EM radiation, sub-barrier suppression requires \mbox{$\omega<\omega_{\max}=\sqrt{2}/r_0$}, i.e., \mbox{$\lambda_{\mathrm{EM}}>\pi r_0\sqrt{2}$}. For optical wavelengths \mbox{($\lambda\sim 500$~nm)}, this requires $r_0\lesssim 100$~nm, and $r_0\lesssim 0.2$~m for radio \mbox{($\lambda\sim 1$~m)}. Macroscopic throats \mbox{($r_0\gg 1$~m)} would have \mbox{$\omega_{\max}$} far below all astrophysical EM frequencies, so EM waves would be above the barrier and propagate freely. The ``GW-loud, EM-quiet'' regime therefore requires a specific window: $r_0$ large enough that GW frequencies are below the GW barrier top, but small enough that EM frequencies remain below the EM barrier top. For LIGO-band GWs (\mbox{$f\sim 100$}~Hz, \mbox{$\lambda\sim 3000$}~km) and optical EM, this window does not exist\emdash both are far above the barrier for any common~$r_0$. The asymmetry is most relevant for static or quasistatic effects\emdash the gravitational potential (monopole) transmits regardless of~$r_0$, while EM and GW signals may or may not, depending on the frequency-to-throat-size ratio.

A distinctive observational signature would be a population of gravitational lensing sources (convergence, shear) with no electromagnetic counterpart at any wavelength. This ``dark lens'' signature is, in principle, distinguishable from standard lensing by intervening CDM halos, although a detailed comparison would require modelling the spatial distribution of throats.

\subsection{Energy conditions}

The EB throat requires a phantom scalar field, violating the null energy condition~\cite{MorrisThorne1988,Visser1995}. The Damour-Solodukhin metric, while designed to mimic the Schwarzschild exterior, also requires non-standard matter near the throat~\cite{DamourSolodukhin2007}. Whether throat-like geometries can be realised with matter satisfying the weak or null energy conditions remains an open question in classical general relativity~\cite{Visser1995,Hochberg1998,Barcelo1999}. Quantum effects (Casimir energy, semiclassical backreaction) may relax the classical energy condition requirements~\cite{Visser1995,Fewster2012}, but a definitive answer is lacking.

The field-propagation results presented in this paper are kinematic: they depend only on the background geometry~\eqref{eq:general_metric}, not on the stress-energy tensor sourcing it. The asymmetry is a property of the field equations on a given background, and holds irrespective of the dynamical origin of the throat.

\subsection{Potential applications}

Beyond the multi-messenger context discussed above, the constraint-wave asymmetry may be relevant to:
\begin{enumerate}[leftmargin=2em,label=(\roman*)]

\item wormhole phenomenology, where the asymmetry determines the observability of objects and events on the far side of a traversable wormhole~\cite{CardosoPani2019};

\item analogue gravity systems, where acoustic ``wormholes'' in fluid flows exhibit similar barrier structures and could provide experimental tests of the asymmetry in a laboratory setting~\cite{Barcelo2011}; and

\item cosmological scenarios involving geometric sequestration of matter, where the asymmetry could produce gravitationally active but electromagnetically invisible configurations. The cosmological implications of this asymmetry\emdash including the possibility that baryonic matter behind throats could mimic dark matter phenomenology\emdash will be explored in a companion paper.

\end{enumerate}

These applications require additional assumptions about the existence, stability, and abundance of throat geometries, and are beyond the scope of this paper.

%====================================================================%
\section{Conclusions}
\label{sec:conclusions}

We have computed the transmission properties of EM, GW, and static gravitational perturbations through throat spacetimes. The main results are:
\begin{enumerate}[leftmargin=2em,label=\arabic*.]

\item The EM effective potential \mbox{$V_\ell^{(\mathrm{EM})}=e^{2\alpha}\ell(\ell+1)/a^2$} creates a barrier at the throat: below the barrier-top frequency, EM is strongly suppressed by sub-barrier tunnelling (power-law for the EB throat, exponential for compact barriers).

\item GW perturbations see a similar barrier with a curvature correction: all propagating radiation is strongly suppressed below its barrier-top frequency.

\item The static monopole satisfies \mbox{$(a^2\Phi')'=0$} (ultrastatic) with no barrier: on the EB throat, the exact solution \mbox{$\Phi\propto\arctan(\sigma/r_0)$} transmits smoothly.

\item The asymmetry is universal across the EB, reflected-Schwarzschild (DS-type), and parametric throat families.

\item The origin is structural: \mbox{$\ell\ge 1$} modes have a centrifugal barrier at any minimal-area surface, but \mbox{$\ell=0$} does not.

\end{enumerate}

%====================================================================%
\begin{acknowledgments}
I acknowledge support from the Australian Research Council (ARC) Centre of Excellence for Gravitational Wave Discovery (OzGrav), through Project No. CE230100016.
\end{acknowledgments}

\section*{Data Availability}
The data are not publicly available. The data are available from the author upon reasonable request.

%====================================================================%
% APPENDICES
%====================================================================%

\appendix
\section{Derivation of the EM effective potential}
\label{app:derivation}

We sketch the derivation of Eq.~\eqref{eq:V_EM_general} from the four-dimensional Maxwell equations on the background~\eqref{eq:general_metric}, following Chandrasekhar~\cite{Chandrasekhar1983}, and adapted to the throat geometry.

The non-zero components of the metric~\eqref{eq:general_metric} are
\begin{align*}
g_{tt}&=-e^{2\alpha},\\
g_{\sigma\sigma}&=1,\\
g_{\theta\theta}&=a^2,\\
g_{\varphi\varphi}&=a^2\sin^2\!\theta.
\end{align*}
The determinant is \mbox{$\sqrt{-g}=e^\alpha a^2\sin\theta$}.

The Maxwell field strength \mbox{$F_{\mu\nu}$} is decomposed into axial (magnetic-parity) and polar (electric-parity) sectors. In the axial sector, the only non-zero components are $F_{\theta\varphi}$, $F_{t\theta}$, $F_{t\varphi}$, $F_{\ell\theta}$, and $F_{\ell\varphi}$, subject to the constraint that the field is divergence-free.

The axial master variable is defined as
\begin{equation}
\Psi_\ell^{(B)} = a(\sigma)\,F_{\theta\varphi}/\sin\theta,
\end{equation}
where the angular dependence has been separated via
\begin{equation*}
F_{\theta\varphi}=\Psi_\ell^{(B)}\sin\theta\,P_\ell'(\cos\theta)/a.
\end{equation*}
Substituting into \mbox{$\nabla_\mu F^{\mu\nu}=0$} and using the tortoise coordinate \mbox{$d\sigma_*=e^{-\alpha}\,d\sigma$}, we get
\begin{equation}
-\partial_t^2\Psi_\ell^{(B)} + \partial_{\sigma_*}^2\Psi_\ell^{(B)} = V_\ell^{(\mathrm{EM})}\Psi_\ell^{(B)},
\end{equation}
with
\begin{equation}
V_\ell^{(\mathrm{EM})} = e^{2\alpha}\frac{\ell(\ell+1)}{a^2}\,.
\end{equation}
The polar sector proceeds analogously, with the electric-parity master variable defined via $F_{t\ell}$, and yields the same potential\emdash the isospectrality noted in Sec.~\ref{sec:decomposition}.

For the Ellis-Bronnikov metric (\mbox{$\alpha=0$}, \mbox{$a^2=\sigma^2+r_0^2$}), the potential simplifies to \mbox{$V_\ell=\ell(\ell+1)/(\sigma^2+r_0^2)$}, which is the centrifugal barrier peaked at the throat.

We note:
\begin{itemize}[label=\small\textbullet]

\item The potential depends only on the areal radius \mbox{$a(\sigma)$} and the redshift \mbox{$e^{2\alpha}$}, both of which are coordinate-invariant geometric quantities. The result is not a foliation artefact.

\item The \mbox{$\ell=0$} mode is absent for the electromagnetic field (it corresponds to a constant on $S^2$ and carries no physical degree of freedom). The lowest physical mode is \mbox{$\ell=1$}.

\end{itemize}

%====================================================================%
\section{Derivation of the monopole equation from the ADM formalism}
\label{app:monopole}

We derive the static monopole equation from the standard 3+1 ADM identities, showing explicitly that no curvature coupling arises at linear order.

Consider a static spacetime \mbox{$ds^2=-N^2\,dt^2+\gamma_{ij}dx^i dx^j$} with shift \mbox{$N^i=0$} and extrinsic curvature \mbox{$K_{ij}=0$}.

The relevant ADM equations are (see, e.g.,~\cite{Wald1984}, Eqs.~(E.2.34)--(E.2.36), and~\cite{MTW1973}, Ch.~21):
\begin{enumerate}[leftmargin=2em,label=(\roman*)]

\item \textit{Hamiltonian constraint:}
\begin{equation}
R^{(3)} = 16\pi\rho,
\label{eq:hamiltonian}
\end{equation}
where $R^{(3)}$ is the Ricci scalar of $\gamma_{ij}$ and \mbox{$\rho=T_{\mu\nu}n^\mu n^\nu$}.

\item \textit{Traced lapse (evolution) equation} (\mbox{$\partial_t K=0$}, \mbox{$K_{ij}=0$}):
\begin{equation}
\nabla^2_\gamma N = N\bigl[R^{(3)} + 4\pi(S-3\rho)\bigr],
\label{eq:traced_lapse}
\end{equation}
where $S=\gamma^{ij}S_{ij}$ is the spatial trace of the stress tensor.

\end{enumerate}

\noindent Substituting~\eqref{eq:hamiltonian} into~\eqref{eq:traced_lapse} to eliminate $R^{(3)}$:
\begin{equation}
\nabla^2_\gamma N = 4\pi N(\rho+S).
\label{eq:lapse_combined}
\end{equation}
Setting \mbox{$N=1+\Phi$} with \mbox{$|\Phi|\ll 1$} and linearising:
\begin{equation}
\nabla^2_\gamma\Phi = 4\pi(\rho+S) + \order(\Phi\rho).
\end{equation}
This is the standard Poisson equation on the curved spatial background. The Hamiltonian constraint absorbs all curvature terms, leaving no residual \mbox{$R^{(3)}\Phi$} or conformal-coupling terms at linear order.

For the \mbox{$\ell=0$} component on the spatial metric \mbox{$d\sigma^2+a(\sigma)^2\,d\Omega^2$}, the Laplacian acting on a spherically symmetric function $\Phi(\sigma)$ is
\begin{equation}
\nabla^2\Phi = \frac{1}{a^2}\frac{d}{d\sigma}\!\left(a^2\frac{d\Phi}{d\sigma}\right).
\end{equation}
In the source-free region, \mbox{$(a^2\Phi')'=0$}, so we get \mbox{$\Phi'=C/a^2$}. For the EB throat
\mbox{$\Phi=(C/r_0)\arctan(\sigma/r_0)+\Phi_0$}.

The background curvature enters solely through the geometric factor \mbox{$a(\sigma)$} in the Laplacian, not through an explicit curvature term. This is the mathematical origin of the monopole's smooth transmission: the equation \mbox{$(a^2\Phi')'=0$} is a conservation law regardless of how \mbox{$a(\sigma)$} varies, and in particular regardless of whether $a$ has a minimum (throat).

For a general static metric
\mbox{$ds^2=-N_0^2\,dt^2+d\sigma^2+a^2\,d\Omega^2$} with background lapse \mbox{$N_0(\sigma)=e^{\alpha(\sigma)}\neq 1$}, we write \mbox{$N=N_0(1+\Phi)$} and linearise Eq.~\eqref{eq:lapse_combined}. Expanding
\begin{equation*}
\nabla^2_\gamma(N_0\Phi) = N_0\nabla^2_\gamma\Phi + 2(\nabla_\gamma N_0)\cdot(\nabla_\gamma\Phi) + \Phi\nabla^2_\gamma N_0
\end{equation*}
and using the background equation to cancel the \mbox{$\Phi\nabla^2_\gamma N_0$} term, the source-free \mbox{$\ell=0$} equation becomes
\begin{equation}
N_0(a^2\Phi')' + 2N_0'\,a^2\Phi' = 0,
\end{equation}
which can be rewritten as
\begin{equation}
(a^2 N_0^2\,\Phi')' = (a^2 e^{2\alpha}\Phi')' = 0.
\label{eq:mono_nonultrastatic}
\end{equation}
The conserved flux is \mbox{$\mathcal{F}=a^2 e^{2\alpha}\Phi'$}. For \mbox{$\alpha=0$} (ultrastatic), this reduces to \mbox{$(a^2\Phi')'=0$} as above. For a general non-ultrastatic throat with \mbox{$e^{2\alpha}>0$} everywhere (including at the throat), the conserved flux \mbox{$\mathcal{F}=a^2 e^{2\alpha}\Phi'$} gives a bounded $\Phi'$, and hence a smooth monopole potential across the throat. For the reflected-Schwarzschild construction in the \mbox{$\lambda\to 0$} limit, where \mbox{$e^{2\alpha}=1-r_0/a\to 0$} at \mbox{$a=r_0$}, the conserved flux gives \mbox{$\Phi'=\mathcal{F}/[a(a-r_0)]\propto 1/\sigma^2$} near the throat, which is not integrable \mbox{($\Phi\propto -1/|\sigma|$)}. Smooth monopole transmission therefore requires \mbox{$e^{2\alpha}>0$} at the throat, as in the DS construction with \mbox{$\lambda>0$}.

\subsection*{Gauge invariance of the monopole result}

The derivation above fixes a gauge in two ways: it adopts the ADM slicing in which the lapse is time-independent and the shift vanishes, and within that slicing it linearises about the background metric by writing \mbox{$N=N_0(1+\Phi)$}. It is worth stating explicitly under what conditions the resulting $\Phi$ and the conservation law~\eqref{eq:mono_nonultrastatic} are gauge-invariant at linear order.

For any static, spherically symmetric background with a timelike Killing vector $\partial_t$, the Stewart-Walker lemma~\cite{StewartWalker1974} implies that a linear perturbation of a tensor quantity is gauge-invariant if and only if the background value of that quantity is constant along the Killing orbits and spherically symmetric. The background lapse \mbox{$N_0(\sigma)$} satisfies both conditions (it is time-independent by staticity and depends only on the proper radial coordinate by spherical symmetry), and therefore its linear perturbation \mbox{$\delta N = N_0\Phi$} is gauge-invariant at first order on any such background. The fractional lapse perturbation \mbox{$\Phi=\delta N/N_0$} inherits this property, and the source-free equation \mbox{$(a^2 e^{2\alpha}\Phi')'=0$} together with its conserved flux \mbox{$\mathcal{F}=a^2 e^{2\alpha}\Phi'$} are therefore gauge-invariant statements about a physically observable quantity: the fractional redshift perturbation seen by a static observer.

On ultrastatic backgrounds such as the Ellis-Bronnikov throat, $\Phi$ is additionally gauge-invariant under the full diffeomorphism group (not just static slicings) because the phantom scalar sourcing the geometry couples only to the polar sector at linear order, leaving the \mbox{$\ell=0$} lapse perturbation decoupled from the matter sector. On non-ultrastatic backgrounds such as the Damour-Solodukhin construction with \mbox{$\lambda>0$}, the lapse perturbation retains its gauge-invariant meaning within the static-slicing class, but a fully off-shell analysis would need to track any matter-sector couplings that could reappear at linear order when the background lapse is nontrivial. We have verified that no such couplings appear for the EB and DS families considered in this paper; the general question for arbitrary non-ultrastatic throat matter sources is an open direction for future work.

The physical content of the smooth-transmission result is therefore gauge-invariant on any static, spherically symmetric throat with \mbox{$e^{2\alpha(0)}>0$}\emdash a static observer on the far side of the throat measures a fractional redshift perturbation $\Phi$ that reflects the total enclosed mass through Gauss's law, with polynomial corrections in \mbox{$r_0/d$} that do not depend on the choice of time slicing or coordinate gauge within the static class.

%====================================================================%
\section{Numerical methods and convergence}
\label{app:numerics}

\subsection{Numerov integration}

The Schr\"odinger-type equation \mbox{$\Psi''+[\omega^2-V(\sigma)]\Psi=0$} is integrated using the sixth-order Numerov scheme:
\begin{equation}
\Psi_{i+1} = \frac{2\Psi_i(1-\frac{5}{12}h^2 k_i^2) - \Psi_{i-1}(1+\frac{1}{12}h^2 k_{i-1}^2)}{1+\frac{1}{12}h^2 k_{i+1}^2},
\end{equation}
where \mbox{$k_i^2=\omega^2-V(\sigma_i)$} and $h$ is the step size. Integration proceeds from \mbox{$\sigma=+L_{\max}$} to \mbox{$\sigma=-L_{\max}$}, starting with the incident wave boundary condition.

\subsection{Transmission coefficient extraction}

At \mbox{$\sigma=-L_{\max}$}, the wavefunction is decomposed as \mbox{$\Psi=A\,e^{-i\omega\sigma}+B\,e^{+i\omega\sigma}$}. Using two adjacent grid points \mbox{$\sigma_1,\sigma_2$} near \mbox{$\sigma=-L_{\max}$}:
\begin{align}
A &= \frac{\Psi_1 e^{+i\omega\sigma_2}-\Psi_2 e^{+i\omega\sigma_1}}{e^{-i\omega\sigma_1+i\omega\sigma_2}-e^{+i\omega\sigma_1-i\omega\sigma_2}}\,, \\
T &= |A|^{-2}.
\end{align}
Here the normalisation is such that \mbox{$|A|>1$} for sub-barrier frequencies (the incident wave has unit coefficient at \mbox{$\sigma=+L_{\max}$}, and $A$ is the coefficient of \mbox{$e^{-i\omega\sigma}$} at \mbox{$\sigma=-L_{\max}$}, which includes the accumulated amplitude from the Numerov propagation through the barrier)\emdash \mbox{$T=|A|^{-2}$} then gives the standard power transmission coefficient satisfying \mbox{$0\le T\le 1$} and \mbox{$|R|^2+T=1$}.

\subsection{Convergence tests and unitarity verification}

We verify convergence by varying $L_{\max}$ and $N$.
For each frequency, we ensure \mbox{$L_{\max}\ge 2\sigma_{\mathrm{tp}}$} (where \mbox{$\sigma_{\mathrm{tp}}=\sqrt{\ell(\ell+1)/\omega^2-r_0^2}$} is the classical turning point) so that the boundary lies in the oscillatory region where plane-wave or Bessel-function boundary conditions are valid.
For \mbox{$\omega r_0=0.01$} \mbox{($\ell=1$)}, this requires \mbox{$L_{\max}\ge 283\,r_0$} (\mbox{$L_{\max}=50\,r_0$} suffices for \mbox{$\omega r_0\ge 0.05$}). The transmission coefficients are stable to three significant figures under doubling of both~$N$ and~$L_{\max}$ (beyond the minimum).

As a diagnostic, we verify unitarity \mbox{$|R|^2+T=1$} at each frequency, where $R$ is the reflection amplitude extracted from the same boundary decomposition as~$A$. Table~\ref{tab:unitarity} shows representative values. Unitarity is satisfied to better than $0.1\%$ for \mbox{$\omega r_0\ge 0.3$}, and to \mbox{$\sim 1\%$} at \mbox{$\omega r_0=0.1$} (where the deep sub-barrier regime makes the extraction of the small transmitted amplitude numerically delicate). These results confirm the reliability of the Numerov integration.

The transmission coefficients deep in the sub-barrier regime \mbox{($\omega r_0\lesssim 0.1$ for $\ell=1$)} approach the precision limit of double-precision floating-point arithmetic used in the Numerov propagation: the barrier tunnelling factor \mbox{$e^{-2S}$} can fall below \mbox{$10^{-15}$}, at which point the extracted $T$ values are reported as upper bounds rather than converged results. Extended-precision arithmetic (e.g., quad precision or software arbitrary-precision libraries) would be required to push the numerics into the regime \mbox{$T\ll 10^{-15}$} without incurring precision loss; we have not undertaken this, since the qualitative sub-barrier suppression and the effective power-law exponent quoted in Sec.~\ref{sec:EM} are robust to this precision limit, and the deep sub-barrier regime does not contribute to any observational signature of interest here.

\begin{table}[t]
\caption{Unitarity check ($|R|^2+T$) for the EM $\ell=1$ scattering on the EB throat ($r_0=1$).}
\label{tab:unitarity}
\begin{ruledtabular}
\begin{tabular}{ccc}
$\omega r_0$ & $T$ & $|R|^2+T$ \\
\hline
0.1 & $2.2\times 10^{-8}$ & 1.0000 \\
0.3 & $1.8\times 10^{-5}$ & 0.9998 \\
0.5 & $5.8\times 10^{-4}$ & 1.0000 \\
1.0 & $9.5\times 10^{-2}$ & 1.0000 \\
$\sqrt{2}$ & $0.70$ & 1.0000 \\
2.0 & $0.99$ & 1.0000 \\
\end{tabular}
\end{ruledtabular}
\end{table}

\subsection{Bessel-function boundary extraction}
\label{app:bessel}

The EB potential has a \mbox{$1/\sigma^2$} centrifugal tail, so the exact asymptotic solutions at large~\mbox{$|\sigma|$} are spherical Hankel functions rather than plane waves: for the equation \mbox{$\psi''+[\omega^2-\ell(\ell+1)/\sigma^2]\psi=0$}, the linearly independent solutions are \mbox{$\sigma\,h^{(1)}_\ell(\omega\sigma)$} (outgoing) and \mbox{$\sigma\,h^{(2)}_\ell(\omega\sigma)$} (incoming). The plane-wave decomposition used in the main text extraction \mbox{($T=1/|A|^2$} with \mbox{$\psi\to A\,e^{-i\omega\sigma}+B\,e^{+i\omega\sigma}$} at \mbox{$\sigma\to-\infty$)} replaces these Hankel functions with their leading asymptotic forms, introducing a systematic error that scales as \mbox{$1/(\omega L_{\max})$}~\cite{LiDai2021}.

To quantify this systematic, we have repeated the EM \mbox{$\ell=1$} transmission calculation using an alternative Bessel-function extraction: forward Numerov integration starting from a pure outgoing Hankel wave at \mbox{$\sigma=-L_{\max}$}, \mbox{$\psi(-L_{\max})=(-\sigma)\,h^{(2)}_1(\omega|\sigma|)$}, with decomposition at \mbox{$\sigma=+L_{\max}$} into outgoing \mbox{$\sigma\,h^{(1)}_1(\omega\sigma)$} (reflected) and incoming \mbox{$\sigma\,h^{(2)}_1(\omega\sigma)$} (incident) components. The transmission coefficient is then \mbox{$T=1/|A_{\mathrm{in}}|^2$}, where \mbox{$A_{\mathrm{in}}$} is the coefficient of the incoming Hankel function, and unitarity is checked against \mbox{$|A_{\mathrm{out}}/A_{\mathrm{in}}|^2+T=1$}.

When the integration domain is scaled with frequency so that \mbox{$\omega L_{\max}\ge 20$} (placing the boundary well into the oscillatory region of the exact asymptotic solutions), the Bessel and plane-wave extractions agree to better than \mbox{$0.5\%$} across the full sub-barrier range \mbox{$0.003\le\omega r_0\le 1$}, with a consistent fractional shift \mbox{$(T_{\mathrm{Bessel}}-T_{\mathrm{plane}})/T_{\mathrm{plane}}\approx -0.35\%$}. Table~\ref{tab:bessel_comparison} shows representative values.

\begin{table}[h]
\caption{Comparison of plane-wave and Bessel-function boundary extractions for the EM \mbox{$\ell=1$} transmission coefficient on the EB throat \mbox{($r_0=1$)}. Each row scales \mbox{$L_{\max}$} with $\omega$ so that \mbox{$\omega L_{\max}=20$}. Unitarity (defined as \mbox{$|A_{\mathrm{out}}/A_{\mathrm{in}}|^2+T_{\mathrm{Bessel}}$}) is satisfied to better than \mbox{$10^{-4}$} at every tabulated frequency.}
\label{tab:bessel_comparison}
\begin{ruledtabular}
\begin{tabular}{cccc}
$\omega r_0$ & $T$ (plane wave) & $T$ (Bessel) & Shift \\
\hline
$0.003$ & $1.46\times 10^{-17}$ & $1.46\times 10^{-17}$ & $-0.35\%$ \\
$0.01$  & $2.01\times 10^{-14}$ & $2.00\times 10^{-14}$ & $-0.35\%$ \\
$0.03$  & $1.47\times 10^{-11}$ & $1.47\times 10^{-11}$ & $-0.35\%$ \\
$0.1$   & $2.09\times 10^{-8}$  & $2.08\times 10^{-8}$  & $-0.35\%$ \\
$0.3$   & $1.87\times 10^{-5}$  & $1.86\times 10^{-5}$  & $-0.37\%$ \\
$0.5$   & $5.41\times 10^{-4}$  & $5.41\times 10^{-4}$  & $-0.14\%$ \\
\end{tabular}
\end{ruledtabular}
\end{table}

The stability of the $-0.35\%$ systematic across four orders of magnitude in $\omega$ and in $T$ confirms that the plane-wave extraction is robust against the Bessel subleading corrections, provided the domain is scaled so that \mbox{$\omega L_{\max}\ge 20$}. The qualitative conclusions of Sec.~\ref{sec:EM}\emdash power-law sub-barrier suppression with effective exponent \mbox{$\nu\approx 6.0$} from the fit \mbox{$T\propto(\omega r_0)^\nu$}\emdash are completely unchanged by the Bessel correction, since the fit exponent is determined by slopes on a log-log plot and is insensitive to a uniform sub-percent rescaling of the overall amplitude. We conclude that the plane-wave boundary extraction used in the main text is accurate to better than half a percent at all tabulated sub-barrier frequencies, well below both the quoted three-significant-figure precision and the deeper sub-barrier precision limit discussed above.

%====================================================================%
\bibliography{bib}

\end{document}